\documentclass[aps,prx,groupedaddress,  superscriptaddress, twocolumn, notitlepage]{revtex4-1}
\usepackage[utf8]{inputenc}
\usepackage{amsmath}
\usepackage{amsfonts}
\usepackage{amssymb}
\usepackage{braket}
\usepackage{graphicx}
\usepackage{booktabs}
\usepackage[colorlinks=true,linkcolor=blue, urlcolor=blue, citecolor=blue]{hyperref}

\newcommand{\normord}[1]{{:}\!\mathrel{#1}\!{:}}
\begin{document}
\title{Model wavefunctions for interfaces between lattice Laughlin states}
\author{B\l a\.{z}ej Jaworowski}
\affiliation{Max-Planck-Institut f\"{u}r Physik komplexer Systeme, D-01187 Dresden, Germany}
\author{Anne E. B. Nielsen}
\altaffiliation{On leave from Department of Physics and Astronomy, Aarhus University, DK-8000 Aarhus C, Denmark}
\affiliation{Max-Planck-Institut f\"{u}r Physik komplexer Systeme, D-01187 Dresden, Germany}

\begin{abstract}
We study the interfaces between lattice Laughlin states at different fillings. We propose a class of model wavefunctions for such systems constructed using conformal field theory. We find a nontrivial form of charge conservation at the interface, similar to the one encountered in the field theory works from the literature. Using Monte Carlo methods, we evaluate the correlation function and entanglement entropy at the border. Furthermore, we construct the wavefunction for quasihole excitations and evaluate their mutual statistics with respect to quasiholes originating at the same or the other side of the interface. We show that some of these excitations lose their anyonic statistics when crossing the interface, which can be interpreted as impermeability of the interface to these anyons. Contrary to most of the previous works on interfaces between topological orders, our approach is microscopic, allowing for a direct simulation of e.g. an anyon crossing the interface. Even though we determine the properties of the wavefunction numerically, the closed-form expressions allow us to study systems too large to be simulated by exact diagonalization.
\end{abstract}
\maketitle
\section{Introduction}
One of the most striking characteristics of topological orders is the bulk-boundary correspondence -- the fact that the bulk properties of the given phase can be inferred from its physics at the edge. This is, however, not a one-to-one relation, as a given bulk phase can have several different kinds of edges even if it is terminated by vacuum \cite{bravyi1998quantum,haldane1995stability,levin2013protected, cano2014bulk}. The edge is therefore richer than the bulk.

Even richer is the physics of interfaces between different topological orders. The investigation of such systems gained significant attention \cite{grosfeld2009nonabelian,gils2009collective,bais2009theory,bais2009condensate, bais2010topological,Beigi2011,kapustin2011topological,Kitaev2012, barkeshli2013classification,
levin2013protected,barkeshli2013theory,lan2015gapped, hung2015generalized,lan2015gapped,cano2015interactions,  barkeshli2015particle,wan2016striped,
santos2017parafermionic,
santos2018symmetry, fliss2017interface,
wang2018topological,
mross2018theory,maymann2019families,crepel2019microscopic, crepel2019model}.
 For example, several authors studied the conditions under which the interfaces are gapped or gapless \cite{Kitaev2012, levin2013protected,maymann2019families, cano2015interactions,fliss2017interface, hung2015generalized,lan2015gapped, barkeshli2013classification, barkeshli2013theory}.  Other works studied the charge and spin of the interface modes \cite{crepel2019microscopic, crepel2019model, grosfeld2009nonabelian}. There has been also a considerable effort dedicated to determining the entanglement entropy and entanglement spectrum at interfaces, which show that interfaces themselves can have a topological structure \cite{crepel2019microscopic, crepel2019model, santos2018symmetry, cano2015interactions, bais2010topological, fliss2017interface,sakai2008entanglement}. All these properties are related to the behavior of fractionalized anyonic excitations at the boundary. Morevover, the physics of gapped interfaces of Abelian states has a close relation to the physics of gapped edges \cite{haldane1995stability,levin2009fractional, neupert2011fractional, cheng2012superconducting, motruk2013topological,  vaezi2013fractional,levin2013protected, cano2014bulk, hung2015ground, kapustin2014ground, wang2015boundary, repellin2018numerical}, as well as the theory of twist defects (genons) \cite{bombin2010twist,barkeshil2013twist, barkeshli2012topological,you2012projective} -- these three can be described within one formalism \cite{barkeshli2013theory, barkeshli2013classification}. In analogy to genons, interfaces of Abelian states can host non-Abelian parafermion zero modes \cite{fendley2012parafermionic, lindner2012fractionalizing,clarke2013exotic, barkeshli2013theory, barkeshli2013classification,klinovaja2014kramers, santos2017parafermionic,wu2018formation,santos2018symmetry, santos2020parafermions}, which have potential applicatons in quantum computing \cite{hutter2016quantum,clarke2013exotic,mong2014universal,lindner2012fractionalizing}.

A particularly relevant class of topologically ordered states are the fractional quantum Hall (FQH) states, which can be created experimentally in a 2D electron gas in a high magnetic field. Interfaces between such states can be created experimentally, when a different filling factor of Landau levels is achieved in different parts of the system \cite{chang1989transmission,camino2005realization}. For example, in an attempt to prove the existence of anyons, an interferometer was created, in which $\nu=2/5$ and $\nu=1/3$ FQH states were placed next to each other \cite{camino2005realization}. The effective theory of the interface, allowing for $e/15$ quasiparticle charge, was invoked in the theoretical description of this experiment \cite{fiete2007universal}. More experiments were proposed to study further kinds of interfaces, e.g. between $\nu=2/3$ spin-polarized domains \cite{wu2018formation}, in graphene \cite{crepel2019microscopic,crepel2019model} or in double quantum wells \cite{yang2017interface}.

In addition to the continuum 2D electron gas, there are also alternative experimental settings for FQH states, some of which are lattice systems. The lattice FQH states can appear in the form of fractional Chern insulators \cite{SunNature,Neupert,PRX}. The presence of the lattice affects the non-universal aspects of FQH physics, and allows for generalizations of the FQH states \cite{SunNature,Neupert,PRX,barkeshli2012topological,liu2012fractional,sterdyniak2013series,
moller2015fractional}. Fractional Chern insulators were created experimentally in Moir\'e lattices in bilayer graphene in a magnetic field \cite{spanton2018observation}. There are also numerous proposals for realizing them in optical lattices \cite{sorensen2005fractional, palmer2006high,palmer2008optical,kapit2010exact,moller2009composite,hafezi2007fractional,yao2013realizing,cooper2013reaching, nielsen2013local}. Such a setting would allow for more control over the system parameters, as well as a realization of the bosonic versions of FQH states, so far not observed experimentally.

The theoretical works on FQH interfaces can be divided into two groups. The first is the “top-down” one, which focuses on field theories and neglects the microscopic details of the systems. The methods involved are based e.g. on $K$-matrices \cite{haldane1995stability,levin2013protected, santos2017parafermionic,fliss2017interface,santos2018symmetry, maymann2019families} or topological symmetry breaking formalism \cite{bais2012modular,bais2009condensate, bais2010topological, bais2009theory, kong2014anyons}  (see also \cite{burnell2018anyon} and references therein). They are powerful tools to determine the universal features of the interfaces, for example they allow for constructing general classifications of the gapped ones \cite{bais2009theory,barkeshli2013theory, barkeshli2013classification,kong2014anyons,hung2015generalized}.

The second approach is the ``bottom-up'' one, less general but more detailed, focused on the microscopic aspects of the system. It provides concrete examples of states belonging to the general classes determined by the ``top-down'' methods, which allows to test the predictions from these works and to investigate the non-universal properties of these states. Such an approach can either rely on diagonalizing Hamiltonians, or on proposing model wavefunctions. Numerical methods of solving the Hamiltonians have size limitations, especially pronounced in the case of interfaces (e.g. due to the reduction of the translational symmetry). Thus, the exact diagonalization \cite{crepel2019microscopic, crepel2019model,liang2019parafermions} or DMRG \cite{zhu2020topological} studies of interfaces are rare and often complemented with other methods. 

Another type of ``bottom-up'' approaches is constructing model wavefunctions. Such constructions are widely used in the case of single quantum Hall states (i.e. without interface) since the seminal work of Laughlin \cite{laughlin}. For such systems, they have proven useful, as they can be studied both analytically and numerically, and for the latter, the considered system sizes can be much larger than in exact diagonalization. They can, but do not have to, be related to Hamiltonians as exact ground states of model Hamiltonians or approximate ground states of short-range ones \cite{haldane1983fractional}. The wavefunctions themselves can provide insights on the nature of certain FQH states (e.g. the mechanism of anyon condensation in the Haldane hierarchy \cite{haldane1983fractional} or composite fermion construction for Jain states \cite{jain1989composite}). A particularly useful way to design model FQH wavefunctions is the conformal field theory (CFT) construction, which has an especially strong link with the ``top-down’’ topological quantum field theories \cite{moore1991nonabelions,hansson2007composite,tu2014lattice, glasser2016lattice,nielsen2018quasielectrons}. 

In the case of interfaces, model wavefunctions were constructed within the matrix product state (MPS) formalism \cite{crepel2019microscopic, crepel2019model, crepel2019variational}. This method builds on infinite-dimensional matrix product states, which can be derived from  CFT for a number of single quantum Hall states \cite{zaletel2012exact, estienne2013matrix,crepel2018matrix,crepel2019matrix}. It was shown that at least in some cases (Halperin-Laughlin, Halperin-Pfaffian), one can connect the matrices belonging to different states and create a wavefunction of the interface \cite{crepel2019microscopic, crepel2019model, crepel2019variational}. Such an approach allows to obtain wavefunctions for the ground state as well as the gapless edge and interface excitations, which were shown to have a high overlap with the exact diagonalization results. The entanglement entropy was studied, confirming the presence of the area law at Halperin-Laughlin interfaces. Although it was not demonstrated explicitly for interfaces, the MPS approach allows also to study bulk quasihole \cite{zaletel2012exact, wu2015matrix} and quasielectron excitations \cite{kjall2018matrix}. 

In this work, we focus on Laughlin-Laughlin interfaces, for which the MPS construction was not yet demonstrated. We construct microscopic model wavefunctions for certain examples of such interfaces in a lattice system. Such a lattice formulation is natural in the context of fractional Chern insulators. We employ a CFT-based method related to, but different from, the one used in Refs.\ \cite{crepel2019microscopic, crepel2019model, crepel2019variational}. Instead of expressing vertex operators of the two CFTs as MPS matrices as in \cite{crepel2019microscopic, crepel2019model, crepel2019variational}, we patch them together directly. In this way, we construct the model wavefunctions for ground state and localized bulk quasihole excitations. Their properties are then studied using Monte Carlo methods \cite{tserkownyak2003monte, barbaran2009numerical,glasser2016lattice,nielsen2018quasielectrons}. With this construction the interface wavefunctions are given in a form resembling the original Laughlin expression \cite{laughlin}. While here we focus on a cylinder geometry with the interface parallel to the periodic direction, like in the MPS works, in general the interface can have any shape, and our wavefunctions are valid for planar systems also. Moreover, our results on quasiholes are easily generalizable to the quasielectrons, which admit a particularly simple description on the lattice \cite{nielsen2018quasielectrons}. 

The paper is organized as follows. In Section \ref{sec:noanyons} we construct the ground state wavefunction, and evaluate its correlation function and entanglement entropy numerically. The former suggests that for short-range Hamiltonians the interface would be gapless (and thus, that it is a different type of interface than studied in Refs.\ \cite{cano2015interactions,maymann2019families, fliss2017interface,santos2017parafermionic, santos2018symmetry}). In Section \ref{sec:anyons} we construct a wavefunction for the quasihole excitations. We perform a microscopic simulation of a quasihole crossing the interface, which was not yet demonstrated. We determine the conditions under which quasihole statistics are well-defined and evaluate the statistical phases. Section \ref{sec:conclusions} concludes the article.

\section{The wavefunctions without anyons}\label{sec:noanyons}
We begin with proposing and studying the model wavefunctions for the ground state of a system with an interface. First (Sec. \ref{ssec:correlator}) we review the CFT construction for a single Laughlin state on the lattice. Next, in Section \ref{ssec:intwfn}, we propose the interface wavefunction, and discuss the conditions in which it is well defined. Because these requirements enforce rather low filling, before presenting concrete examples, we study both sides of the interface separately and show that they are topological (Sec. \ref{ssec:onelaughlin}). Then, in Sections \ref{ssec:correlation} and \ref{ssec:interface_ent}, respectively, we determine numerically the correlation function and entanglement entropy for a system with an interface.

\subsection{Model wavefunctions from CFT - preliminaries}\label{ssec:correlator}

The Laughlin states \cite{laughlin}, occurring at filling factor $\nu=1/q$, $q\in \mathbb{N}^{+}$, of the first Landau level, are the simplest fractional quantum Hall (FQH) states, with the $\nu=1$ integer quantum Hall effect being a special case at $q=1$. Each $q$ corresponds to a different topological order, with excitations, quasielectrons and quasiholes, having fractional charge (a multiple of $e/q$) and fractional statistics (exchange phase being a multiple of $\pi/q$). Here we review the CFT construction of the lattice versions of these states in planar geometry from Refs.\ \cite{tu2014lattice,glasser2016lattice}, which builds on the framework proposed by Moore and Read for continuum FQH states  \cite{moore1991nonabelions}. 

The general form of the lattice wavefunction is 
\begin{equation}
\ket{\psi}=\frac{1}{C}\sum_{\mathbf{n}}\psi(\mathbf{n})\ket{\mathbf{n}}
\label{eq:wfn_general}
\end{equation}
where $\mathbf{n}=[n_1, n_2,\dots, n_N]$ is the vector of occupation numbers of the lattice sites, $\ket{\mathbf{n}}$ is the corresponding Fock-space basis state (we assume that the creation operators in the definition of $\ket{\mathbf{n}}$ are sorted by the site index), $N$ is the number of sites and $C$ is the normalization constant. The wavefunction can describe either fermions or bosons, but we enforce the hard-core condition for the latter, i.e. $n_i\in\{0,1\}$ in both cases. Since we have a discretized system, we consider a magnetic field which penetrates only the lattice sites. We describe it by associating a positive real number $\eta_i$ with each site $i$. This number describes the number of flux quanta passing through that site.

In the CFT construction, the squared modulus of the wavefunction coefficient can be expressed by a correlator of a conformal field theory with compactification radius $\sqrt{q}$, which can be written as
\begin{equation}
|\psi(\mathbf{n})|^2\propto \braket{0|\prod_{i=1}^N V(n_i, z_i, \bar{z}_i)|0}
\label{eq:correlator}
\end{equation}
where $\ket{0}$ is a vacuum of this CFT, $z_i=x_i+iy_i$ is a coordinate of a lattice site $i$, $\bar{z}_i$ is its complex conjugate and  $V(n_i, z_i,  \bar{z}_i)$ is a vertex operator defined by
\begin{equation}
V(n_i, z_i, \bar{z}_i)=\normord{\exp(i\gamma_i(n_i)\phi(z_i,\bar{z}_i))}
\label{eq:vertex}
\end{equation}
where $\phi(z_i,\bar{z}_i)$ is a free bosonic field, and $\gamma_i$ is a function of the occupation of a lattice site $i$, given by $\gamma_i(n_i)=\frac{qn_i-\eta_i}{\sqrt{q}}$. Evaluating the correlator, we arrive at the following expression for the unnormalized wavefunction
\begin{equation}
\psi(\mathbf{n})=\chi_{\mathbf{n}}\delta_{\gamma}\prod_{i<j}(z_i-z_j)^{\gamma_i(n_i)\gamma_j(n_j)}
\label{eq:cft_wfn_general}
\end{equation}
Here, $\delta_{\gamma}=\delta(\sum_{i=1}^{N}\gamma_i(n_i))$ and $\chi_{\mathbf{n}}$ is an arbitrary phase factor. Since the wavefunction is constructed as a product of vertex operators, it is natural to choose a phase factor as a product $\chi_{\mathbf{n}}=\prod_i\chi_i(n_i)$, where $\chi_i(n_i)$ depends only on the occupation of site $i$. Under such an assumption, the quantities we calculate in this Section (particle density, correlation function, entanglement entropy) do not depend on the particular choice of $\chi_i(n_i)$, thus we set $\chi_i(n_i)=1$ without loss of generality.

Substituting the explicit expression for $\gamma_i(n_i)$, and disregarding some factors influencing only the normalization or the gauge, we obtain the following wavefunction coefficients
\begin{equation}
\psi(\mathbf{n})=\delta_{\mathbf{n}}\prod_{i<j}(z_i-z_j)^{qn_in_j}\prod_{i\neq j}(z_i-z_j)^{-n_i\eta_j}
\label{eq:onelaughlin}
\end{equation}
 and $\delta_{\mathbf{n}}=\delta(qM-N_{\phi})$, where $M=\sum_{i=1}^N n_i$ is the total number of particles, and $N_{\phi}=\sum_{i=1}^N \eta_i$ is the number of magnetic flux quanta passing through the system. Thus, $\delta_{\mathbf{n}}$ enforces the charge neutrality (the background charge is included in the vertex operators describing sites, in contrast to the continuum case, where an additional vertex operator for the background charge has to be added \cite{moore1991nonabelions}). Because in general $N_{\phi}\neq N$, the wavefunction \eqref{eq:onelaughlin} can be characterized by two filling factors -- the ``Laughlin filling'' $\nu=M/N_{\phi}=1/q$, defined as the number of particles per magnetic flux quantum, determining the topological class of the wavefunction, and the ``lattice filling'' $\nu_{\mathrm{lat}}=M/N$, defined as the number of particles per site, controlled by $\eta_i$. By tuning $\eta_i$ one can interpolate between continuum and lattice states \cite{glasser2016lattice}, with the last term of \eqref{eq:onelaughlin} becoming the exponential term of the usual Laughlin function for infinite systems in the continuum limit $\eta_i \rightarrow 0$. We note that for certain values of $\gamma$ and $\eta$, one can use CFT to derive a Hamiltonian for which \eqref{eq:onelaughlin} is an exact ground state \cite{tu2014lattice,glasser2016lattice}.

In the following, we will work in the cylinder geometry rather than the planar one. Throughout this work, we will assume that the direction $y$ is a periodic one. Let $L$  be the circumference of the cylinder. Given a set of coordinates $\{\zeta_1, \zeta_2, \dots \zeta_N\}$ on a cylinder, we relate them to the plane coordinates as $z_i=e^{2\pi i \zeta_i/L}$. By substituting the resulting $z_i$ into $\eqref{eq:cft_wfn_general}$, one obtains a wavefunction on a cylinder (see e.g. Ref.\ \cite{glasser2016lattice}).

\begin{figure}
\includegraphics[width=0.5\textwidth]{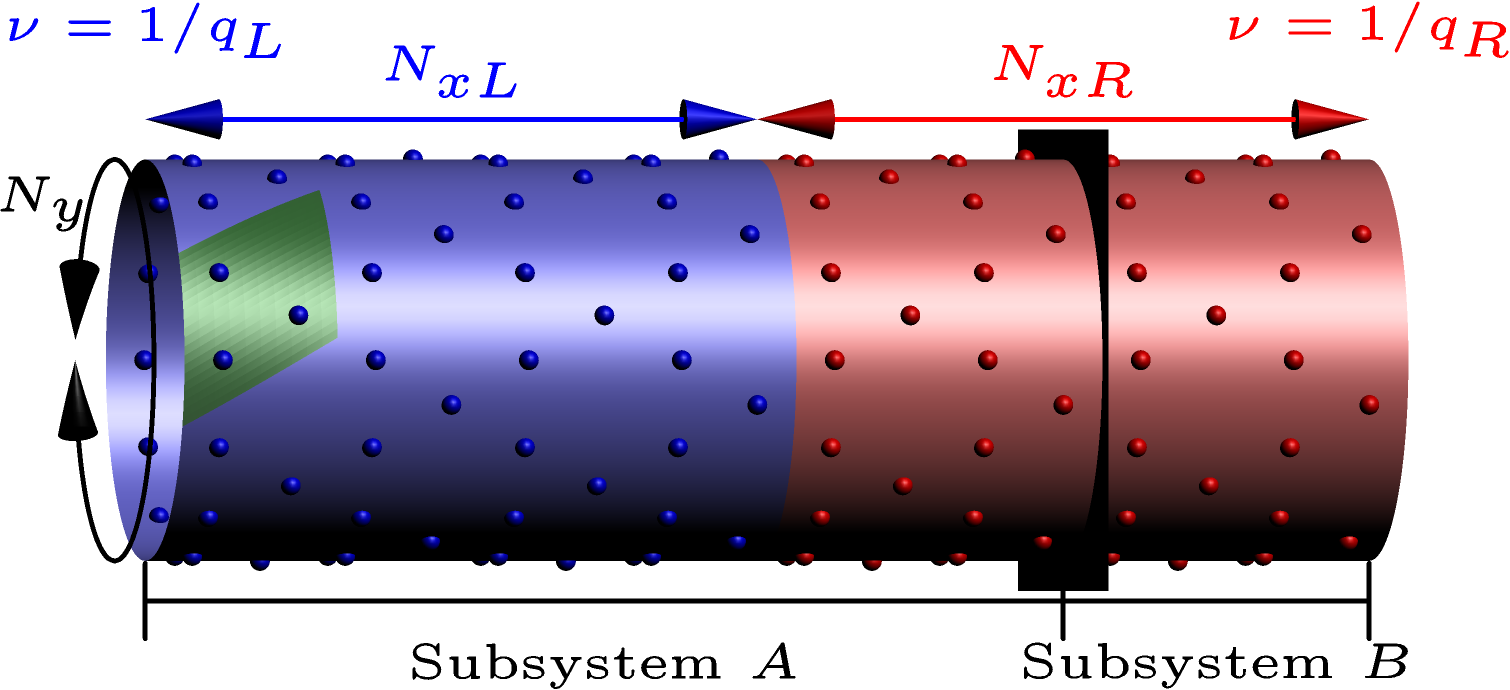}
\caption{The interface considered in this work. Blue and red colors correspond to two Laughlin fillings $\nu=1/q_L$, $\nu=1/q_R$, respectively, with spheres denoting the lattice sites. The green rhombus denotes the unit cell of the kagome lattice. The black plane is an example of the entanglement cut, dividing the cylinder into two cylindrical subsystems $A$, $B$.}
\label{fig:system}
\end{figure}

\subsection{Model wavefunction for the interface}\label{ssec:intwfn}

We proceed to describing an interface between two Laughlin states with different Laughlin filling factors $\nu_I=1/q_I$, where $q_I \in \mathbb{N}^{+}$ and $I\in \{L,R\}$ denotes the left and right sides of the interface, respectively. Such interfaces were studied within the top-down approach. More precisely, these works concentrated on a subset of such systems, namely the gapped interfaces. At filling factors fulfilling
\begin{equation}
q_L a^2=q_R b^2,
\label{eq:filling_condition}
\end{equation}
with $a$, $b\in \mathbb{N}^{+}$, such an interface can be gapped by certain types of interactions breaking particle number conservation (which can in principle be realized by coupling to a superconductor) \cite{cano2015interactions,levin2013protected,maymann2019families, fliss2017interface}. In such a case, the tunneling of the anyons through the interface is restricted (in a simplest case $b=1$, $a$ right-type anyons should tunnel at the same time), which gives rise to nontrivial properties such as anyonic Andreev reflection \cite{santos2017parafermionic}, correction to entanglement entropy scaling \cite{cano2015interactions,santos2018symmetry} or parafermionic modes  \cite{santos2017parafermionic}.

Here we construct microscopic model wavefunctions for Laughlin-Laughlin interfaces by generalizing the CFT construction from Sec.\ \ref{ssec:correlator}. We consider a system where the $L$ and $R$ parts contain $N_L$, $N_R$ sites, respectively (see Fig.\ \ref{fig:system}). The total number of sites is denoted by $N=N_L+N_R$. We denote the coordinates of the sites as $z_i=z_{i;L}$ for $i=1,\dots, N_L$ and $z_i=z_{i-N_L;R}$ for $i=N_L+1,\dots, N$ (analogous indices will be used for their occupation numbers). We are going to construct the wavefunction as a correlator of the vertex operators describing the two Laughlin states at fillings $\nu_I=1/q_I$, $I\in \{L,R\}$. Each vertex operator corresponds to a site of part $I$. We note that the lattice nature of our wavefunctions is crucial to our construction -- since the vertex operators describe sites, they can be easily divided into two sets based on their position. In the continuum, the vertex operators correspond to particles and such a division is less straightforward.

In general, a correlator of the vertex operators belonging to two different CFTs cannot be evaluated. This imposes a restriction on possible filling factors on both sides of the interface \cite{bais2009condensate,bais2010topological} Only when \eqref{eq:filling_condition} is fulfilled, both CFTs can be embedded in a third CFT with compactification radius $a\sqrt{q_L}=b\sqrt{q_R}$, thus the construction \eqref{eq:onelaughlin} is still valid in such a case. Note that this does not mean that the interface is necessarily gapped -- this depends on the interaction generating our wavefunction. We focus on a particular choice $b=1$, in which the left CFT can be embedded in the right one. We note that while our approach describes a certain type of interface at certain filling factors, in reality one can put two Laughlin states with any filling factors next to each other. Thus, to obtain a complete understanding of all possible Laughlin-Laughlin interfaces, our method should be complemented with other methods, such as exact diagonalization or DMRG.

We choose the vertex operators in such a way that the first $N_L$ of them describe a Laughlin state with filling $\nu=1/q_L$ and constant $\eta_i=\eta_L$, while the next $N_R$ correspond to a similar state with filling $\nu=1/q_R$ and constant $\eta_i=\eta_R$. That is, they have the form \eqref{eq:vertex}, with
\begin{equation}
\gamma_i (n_i)=\begin{cases}
\frac{q_{L}n_{i;L}-\eta_L  }{\sqrt{q_L}} & \mathrm{ for }~i=1,\dots, N_L \\
\frac{q_{R}n_{i-N_{L};R}-\eta_R  }{\sqrt{q_R}} & \mathrm{ for }~i=N_L+1,\dots, N
\end{cases}.
\label{eq:intwfn1}
\end{equation}

The result is the following expression for the wavefunction coefficients
\begin{equation}
\psi(\mathbf{n})\propto \delta_{\mathbf{n}_L, \mathbf{n}_R}\psi_L(\mathbf{n}_L)\psi_R(\mathbf{n}_R)\psi_{LR}(\mathbf{n}),
\label{eq:interfacewfn}
\end{equation}
where $\psi_L(\mathbf{n}_L), \psi_R(\mathbf{n}_R)$ are the Laughlin wavefunctions \eqref{eq:onelaughlin} at the respective side of the interface (disregarding the charge neutrality),
\begin{equation}
\psi_L(\mathbf{n}_L)=\prod_{i<j}(z_{i;L}-z_{j;L})^{q_Ln_{i;L}n_{j;L}}\prod_{i\neq j}(z_{i;L}-z_{j;L})^{-n_{i;L}\eta_{L}}
\end{equation}
\begin{equation}
\psi_R(\mathbf{n}_R)=\prod_{i<j}(z_{i;R}-z_{j;R})^{q_Rn_{i;R}n_{j;R}}\prod_{i\neq j}(z_{i;R}-z_{j;R})^{-n_{i;R}\eta_{R}},
\end{equation}
while $\psi_{LR}(\mathbf{n})$ describes the cross factors
\begin{multline}
\psi_{LR}(\mathbf{n})=\prod_{i,j}(z_{i;L}-z_{j;R})^{aq_{L}n_{i;L}n_{j;R}}\times \\ \times
\prod_{i,j}(z_{i;L}-z_{j;R})^{-n_{i;L}\eta_{R}/a-n_{i;R}\eta_{L}a}.
\end{multline}
Note that the mutual statistics of the particles on the two sides of the interface is controlled by $aq_L$ which is always an integer, therefore they are always bosonic or fermionic.

\begin{figure}
\includegraphics[width=0.45\textwidth]{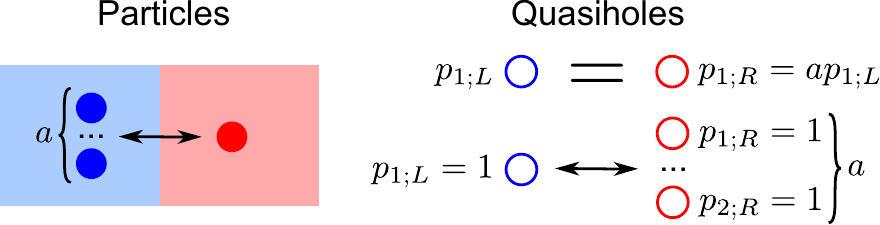}
\caption{Charge conservation in our system. A single $R$-type particle (red filled circle) has the same charge as $a$ $L$-type particles (blue filled circles). On the other hand, an $L$-type quasihole (blue empty circle) has the same charge as e.g. $a$ $R$-type quasiholes with the same $p$ or one $R$ type quasihole with $a$ times larger $p$. In the latter case the two kinds of quasiholes are exactly the same object, which is signified by the ``='' sign. Note that while particles are confined to their ``parent'' part, the quasiholes can be located anywhere. This is emphasized in the figure by placing the circles representing the particles, but not the quasiholes, on the background of the respective color.
}
\label{fig:chargeconservation}
\end{figure}

The charge neutrality is enforced by 
\begin{equation}
\delta_{\mathbf{n}_L, \mathbf{n}_R}=\delta\left( q_L\left(M_L+aM_R\right) -N_{\phi ;L}-N_{\phi ;R}/a\right)
\label{eq:intcn1}
\end{equation}
where $M_{I}=\sum_{i}n_{i;I}$, $N_{\phi;I}=N_{I}\eta_{I}$, $I\in\{L,R \}$. This means that the particle number is not conserved, as destroying one particle on the right means creating $a$ particles on the left (this rule is illustrated in Fig.\ \ref{fig:chargeconservation}, along with an analogous one for quasiholes, which will be derived in Sec. \ref{ssec:anyons_wfn}). Such a behavior may be counterintuitive, but not unexpected -- the  ``top-down'' works predict that precisely this kind of particle number conservation breaking is necessary to gap out the interface (but not sufficient -- it also depends on the interaction Hamiltonian at the interface) \cite{cano2015interactions, santos2017parafermionic}. As stated in Ref.\ \cite{santos2017parafermionic}, this can be interpreted either by assigning the same charge to all the particles, and breaking the charge conservation by coupling the interface to a superconductor, or by assuming that the $R$ particles have $a$ times more charge than the $L$ ones, and retaining the charge conservation. Since the second interpretation would be more convenient later when studying the quasiholes, we fix the charge of $L$, $R$ particles to 1 and $a$, respectively.

The charge neutrality rule \eqref{eq:intcn1} makes the physical realization of our interface challenging. If we consider a realization in solid state (e.g. Moir\'{e} superlattices \cite{spanton2018observation}), then the interaction at the interface has to be mediated by Cooper pairs, i.e. coupling to a superconductor is required (this was already mentioned in Ref.\ \cite{cano2015interactions,santos2017parafermionic}), and fermions on both sides (odd $a$ and $q_L$) need to be considered. The most plausible fermionic case is $q_L=1$, $a=3$, which would require $\nu_R=1/9$, impossible to realize in an ordinary 2D electron gas. Nevertheless, since the interaction in Moir\'e superlattices does not mimic exactly the one in the continuum Landau level, we do not rule out the possibility of observing a $1/9$ FQH state there. 

Otherwise, we may consider e.g. optical lattices \cite{sorensen2005fractional, palmer2006high,palmer2008optical,kapit2010exact,moller2009composite,hafezi2007fractional,yao2013realizing,cooper2013reaching, nielsen2013local}, where one can realize both bosons and fermions. If the optical lattice is interpreted as a spin system, then, in the simplest case $a=2$, one can avoid breaking $S_z$ conservation by representing the $L$ side with $S=1/2$ sites and the $R$ side with $S=1$ sites with strong penalty on $S_z=0$ preventing the spins from being in this state, i.e. $n_i=0,1$ states would be represented by $S=\pm 1/2$ on the left and $S=\pm 1$ on the right.

In general, the wavefunction coefficients \eqref{eq:interfacewfn} are not invariant under the scaling of coordinates $z\rightarrow bz, b\in \mathbb{C}$, in contrast to the Laughlin wavefunction \eqref{eq:onelaughlin}, where the scale is arbitrary. This invariance can be restored by setting the lattice filling to be $\nu_{\mathrm{lat}}=1/2$ at both sides, which can be done by adjusting $\eta_{L,R}=q_{L,R}/2$. We will enforce this condition throughout this work. We note that it is especially suited for spin systems, as then $\frac{q_In_i-\eta_I}{\sqrt{q}}=\sqrt{q}s_i$, with $s_i=\pm 1$.

In this work, we focus only on the wavefunction, without considering the Hamiltonian generating it. However, as we noted before, one can derive the parent Hamiltonians for single lattice Laughlin states for certain values of $q$ and $\eta$. Under the $\eta=q/2$ condition, it is possible to derive $q=1$, $q=2$ and $q=4$ Hamiltonians, therefore in the special case $q_L=1$, $a=2$ we can generate the parent Hamiltonians for both sides of the interface separately \cite{tu2014lattice, glasser2016lattice}. However, because the two Hamiltonians are derived using slightly different methods, they cannot be easily generalized to an interface Hamiltonian. Another way of making connection between our wavefunction and the Hamiltonian may be to look at a general short-range Hamiltonian and optimize the overlap of its ground state with our wavefunction. This was a successful approach for some single lattice quantum Hall states \cite{nielsen2013local, nandy2019truncation}, although we note that due to the size limitations and the shape of the kagome lattice (which we have to choose to ensure correct topological properties, see Sec.\ \ref{ssec:onelaughlin}), making the system  prone to edge effects, the exact diagonalization can have limited applicability. However, for relatively thin cylinders, it may be also possible to replace it with DMRG \cite{grushin2015characterization}.

\begin{figure}
\includegraphics[width=0.5\textwidth]{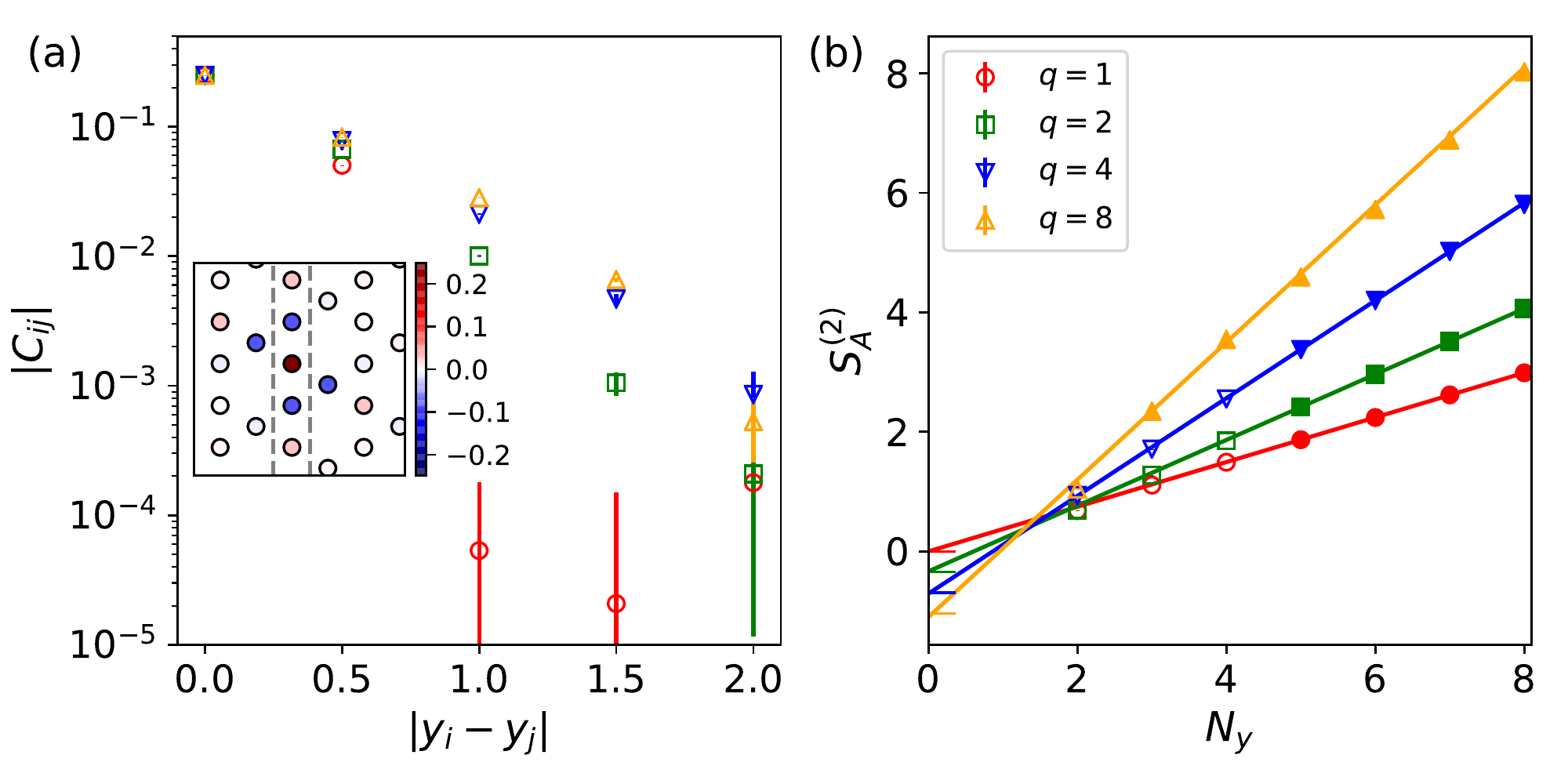}
\caption{Properties of the single Laughlin states considered in this work. (a) The absolute value of the correlation function at constant $x$ in an $N_x\times N_y=8\times 8$ cylinder. The inset shows the spatial profile of the correlation function close to site $i$ (brown circle at the center) for the $q=8$ case. The main plot shows the results for the strip of sites between the dashed gray lines. The values for two sites in the same distance to site $i$ (above and below) are averaged in the main plot. (b) The R\'enyi entropy $S^{(2)}_A$ as a function of the cylinder circumference $N_y$, with the cut at the middle of the sample. The corresponding cylinder length is $N_x=2\lceil N_y/2 \rceil$, with $\lceil~ \rceil$ denoting the ceiling function. Lines denote linear fits.  Only the data points denoted by filled symbols are taken into account in the fitting procedure, because the data points at small $N_y$ are more strongly affected by finite size effects. The regression is weighted based on the Monte Carlo uncertainties. The colorful ticks on the $y$ axis denote the theoretical values $\ln(q)/2$ of the topological entanglement entropy}
\label{fig:onestate}
\end{figure}

\subsection{Single Laughlin wavefunctions - numerical calculations}\label{ssec:onelaughlin}

At $\eta=q/2$ and low filling ($q>4$), the lattice Laughlin states on the square lattice develop long-range antiferromagnetic correlations which destroy the topological order \cite{glasser2016lattice}. To prevent this from happening, we need to work on a frustrated lattice, on which a N\'{e}el ordering is impossible. We choose the kagome lattice, which, according to Ref.\ \cite{glasser2016lattice}, has shorter correlation length than the triangular lattice, and thus smaller finite-size effects. Unless noted otherwise, throughout this work we consider systems on a cylinder, with $(N_{xL}+ N_{xR} )\times N_y$ unit cells, as shown in Fig.\ \ref{fig:system}. For concreteness, we set the lattice constant (i.e. the length of one of the edges of the green rhombus in Fig.\ \ref{fig:system}) to 1, i.e. the nearest-neighbor distance to 0.5, although the wavefunction would not change if the coordinates are rescaled.

To show that we have a topological state on both sides, we first study single Laughlin states ($q_L=q_R=q$, $N_{xL}+N_{xR}=N_x$) before proceeding to the interfaces. We focus on the cases $q=1,2,4,8$, necessary for two examples of $a=2$ interfaces with $q_L=1$ and $q_L=2$. The expectation value of any operator $\hat{O}$ that is diagonal in the occupation number basis $\hat{O}=\sum_{\mathbf{n}}O_{\mathbf{n}} \ket{\mathbf{n}}\bra{\mathbf{n}}$ can be written as
\begin{equation}
\langle \hat{O} \rangle =\frac{\sum_{\mathbf{n}} O_{\mathbf{n}} |\psi(\mathbf{n}) |^2 }{ \sum_{\mathbf{n}} |\psi(\mathbf{n}) |^2 }.
\end{equation}
It can be sampled using Metropolis Monte Carlo, treating $ |\psi(\mathbf{n}) |^2$ as the probability distribution. In such a way, we can evaluate the density-density correlation function
\begin{equation}
C_{i,j}=\langle n_i n_j  \rangle-\langle n_i\rangle \langle  n_j  \rangle
\end{equation}

The results for $\nu=1/4$ and $\nu=1/8$ are shown in Fig \ref{fig:onestate} (a). Although the correlation shows some antiferromagnetic-like behavior at short distances, its absolute value is decaying exponentially, showing the lack of antiferromagnetic ordering.

If exponential decay is assumed, the correlation length may be estimated by the quantity \cite{glasser2016lattice}
\begin{equation}
d_{q}=\frac{1}{2(\ln|C_{i,j}|-\ln|C_{i,k}|))}
\end{equation}
where $i,j,k$ have the same $x$ coordinate, and $|y_i-y_j|=1/2$, $|y_i-y_k|=1$ (note that we set the nearest neighbour distance to 0.5, which generates the factor of 2 in the denominator).  We obtain $d_1=0.07\pm0.3, d_2= 0.264 \pm 0.004, d_4=0.383 \pm 0.008, d_8=0.464 \pm 0.006$.
 
Next, we study the entanglement entropy. We divide the system into two subsystems $A$, $B$ with a cut along the periodic direction of the cylinder (see Fig.\ \ref{fig:system} and consider a special case with just one type of Laughlin states).  We choose the R\'{e}nyi entropy of order 2, $S^{(2)}_A=-\ln \left(\mathrm{Tr} \left(\rho_A^2\right) \right)$, where $\rho_A$ is the reduced density matrix of subsystem $A$. It can be calculated using the Monte Carlo method and the replica trick \cite{cirac2010infinite, hastings2010measuring}. We consider two copies of the system, and write 
\begin{multline}
\mathrm{Tr} \left(\rho_A^2 \right)=\sum_{\mathbf{m},\mathbf{n}}|\psi(\mathbf{m}_A, \mathbf{m}_B)|^2|\psi(\mathbf{n}_A, \mathbf{n}_B)|^2\times \\ \times
\frac{\psi(\mathbf{m}_A, \mathbf{n}_B)\psi(\mathbf{n}_A, \mathbf{m}_B)}{\psi(\mathbf{m}_A, \mathbf{m}_B)\psi(\mathbf{n}_A, \mathbf{n}_B)}
\label{eq:replica}
\end{multline} 
 where $\mathbf{m}_A$, $\mathbf{m}_B$ denote the occupation numbers within the respective subsystems for the first copy of the system, and $\mathbf{n}_A$, $\mathbf{n}_B$, analogically, for the second copy. If the total charge changes after swapping $\mathbf{m}_B\rightarrow \mathbf{n}_B$,  $\mathbf{n}_B\rightarrow \mathbf{m}_B$, then the charge neutrality enforces $\psi(\mathbf{m}_A, \mathbf{n}_B)=\psi(\mathbf{n}_A, \mathbf{m}_B)=0$.  Eq. \eqref{eq:replica} can be evaluated numerically using the Metropolis Monte Carlo method, treating $|\psi(\mathbf{m}_A, \mathbf{m}_B)|^2|\psi(\mathbf{n}_A, \mathbf{n}_B)|^2$ as the probability distribution in importance sampling.

In two-dimensional gapped systems with a single topological phase, the entanglement entropy for a spatial bipartition has a linear scaling (area law) with a constant term,
\begin{equation}
S^{(2)}_A(N_y)=AN_y-\gamma,
\label{eq:fit}
\end{equation}
where $A$ is a nonuniversal coefficient, while the constant term $\gamma$ has the interpretation of topological entanglement entropy, characterizing the given topological order \cite{kitaev2006topological, levin2006detecting}. For our lattice Laughlin states, the entanglement entropy as a function of the cylinder circumference is shown in Fig.\ \ref{fig:onestate} (b).  The results, in general, show the adherence to the area law \eqref{eq:fit}. To obtain $\gamma$, we perform a linear fit with weights based on the Monte Carlo errors. For $q=1,2,4$, after excluding several data points for low circumferences, which are influenced by finite-size effects, we obtain $\gamma$ close to the theoretical prediction $\ln(q)/2$. From the fits we get $\gamma=0.001 \pm 0.007$, $\gamma=0.334 \pm 0.005$,  $\gamma=0.70 \pm 0.03$, close to $\ln(1)/2=0$, $\ln(2)/2\approx 0.346$,  $\ln(4)/2\approx 0.69$ for $q=1,2,4$, respectively (the errors here are the uncertainties of the fit only). For $q=8$ the situation is more complicated. The fit presented in Fig.\ \ref{fig:onestate} (b) yields $\gamma=1.10 \pm 0.07$, a relatively good match with the theoretical value $\ln(8)/2\approx 1.04$. However, the data points from the Monte Carlo calculation exhibit some oscillations around the linear trend. This leads to a strong dependence of the fitted $\gamma$ on the included data points. For example, using only $N_y>3$ we obtain $\gamma=0.81\pm 0.05$, which is further away from the expected value. This can be explained by comparing $N_y$ to the correlation length. The largest investigated system has circumference larger than 100 correlation lengths in the $q=1$ case and less than 22 correlation lengths in the $q=8$ case. Therefore, we can estimate that each dimension of the system needs to be 5 times larger in the $q=8$ case than in the $q=1$ case to obtain the same strength of finite-size effects (which is difficult to achieve within our Monte Carlo approach). Nevertheless, even though for $q=8$ we cannot determine the value of $\gamma$ accurately, the obtained values indicate that it is nonzero, and thus that the state is topological.

\begin{figure}
\begin{center}
\includegraphics[width=0.5\textwidth]{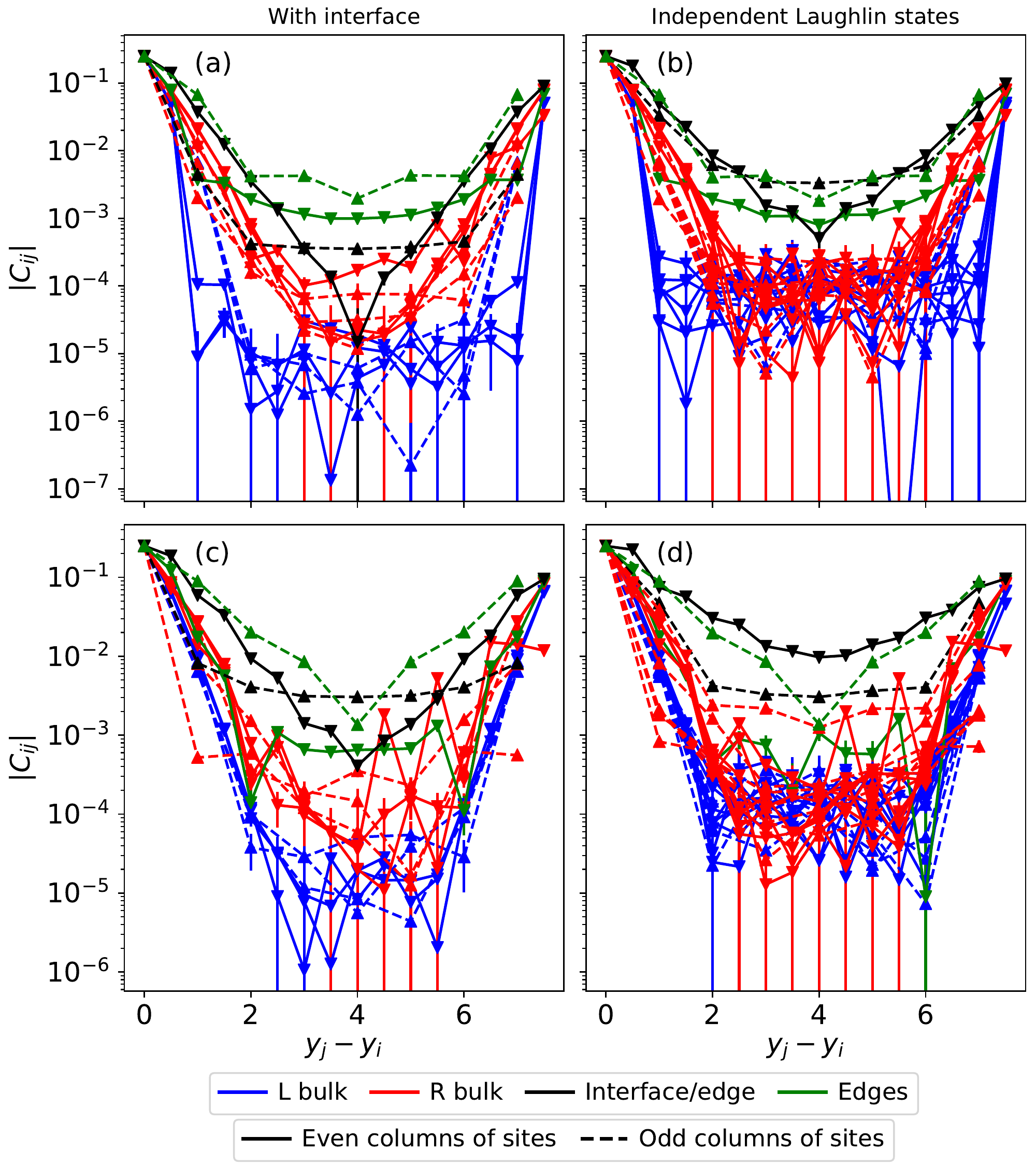}
\end{center}

\caption{Correlation functions for $(4+4)\times 8$ systems with an interface ((a),(c)) compared to superimposed results for the respective single Laughlin states in $8\times 8$ systems ((b),(d)). The upper row refers to the case  $q_L=1$, $a=2$ while the lower one contains results for the case $q_L=2$, $a=2$. The correlation functions are computed along the $y$ direction, each curve corresponding to a different, fixed $x$ coordinate. In (a), (c) the red and blue colors denote the bulks of the $L$ and $R$ parts, in (b), (d) the bulks of the respective single Laughlin states. The solid and dashed lines denote the even and odd columns of sites. The former contain $2N_y$ sites while the latter only $N_y$ sites. In (a), (c) the green and black correspond to the edges and interface (two closest columns of sites), respectively, while in (b), (d) these colors denote the corresponding edges.
}
\label{fig:correlation_interface}
\end{figure}

\subsection{Correlation function at the interface}\label{ssec:correlation}
Next, we study the whole interface wavefunction. We consider two examples, $q_L=1$ and $q_L=2$, both with $a=2$. The former describes an interface between a fermionic integer quantum Hall state and a bosonic $\nu=1/4$ Laughlin state, the latter corresponds to an interface between two bosonic Laughlin states at $\nu=1/2$, $\nu=1/8$. In both cases, the Monte Carlo evaluation of the particle density yields $\braket{n_i}\approx 1/2$ in the entire system, which agrees with the fact that $\eta_I=q_I/2$ correspond to lattice half-filling.

We would also like to determine if the interface is gapped or gapless. This is impossible if we do not put any restriction on the possible parent Hamiltonians (note that lattice parent Hamiltonians generated from CFTs are long-range \cite{tu2014lattice, glasser2016lattice}), as it is always possible to write the Hamiltonian as a sum of projections on its eigenstates $\ket{i}$, $H=\sum_iE_i\ket{i}\bra{i}$. Knowing only the ground state $\ket{0}$, we can choose the energies $E_i$ and other eigenstates $\ket{i}$, $i>0$ arbitrarily (up to the condition that they should be orthogonal to $\ket{0}$ and each other). Nevertheless, for short-range Hamiltonians some indication of the existence of the gap can be obtained from the correlation function -- for nondegenerate ground states of gapped short-range Hamiltonians it typically vanishes exponentially.

 For our system, we evaluate the correlation function as a function of distance between sites in the $y$ direction for different $x$ positions, see Fig.\ \ref{fig:correlation_interface}. In Fig.\ \ref{fig:correlation_interface} (a), showing results for a $q_L=1$, $a=2$ system, it can be seen that in the left and right bulks (blue and red, respectively), the correlation function decays exponentially, until the relative Monte Carlo error gets large. However, on the edges (green lines), it does not, it achieves an approximately constant nonzero value at large enough distances. This is consistent with the fact that the edges of Laughlin states are gapless.
 
 A similar behavior is seen on the rightmost sites of the part $L$ (black dashed line), i.e. next to the interface. We can compare this to a case of two separate quantum Hall states (Fig.\ \ref{fig:correlation_interface} (b)), in which we simply superimpose the result for single Laughlin states with $\nu=1/q_L$ and $\nu=1/q_R$.  We can see that on the rightmost sites of the $L$ part the behavior of the correlation function is similar in the two cases, although the minimum value of the correlation function is smaller for the interface. As for the leftmost sites of the $R$ part, in the interface case the correlation function seems to fall exponentially, while for single Laughlin states it does not. Thus, the correlation function suggests that if the interface can be generated by a short-range Hamiltonian, it is probably gapless, in contrast to the cases studied in Refs.\ \cite{cano2015interactions,maymann2019families, fliss2017interface,santos2017parafermionic, santos2018symmetry}. A similar conclusion can be drawn for $q_L=2$, $a=2$ (Fig \ref{fig:correlation_interface} (c), the results for corresponding single Laughlin states are plotted in \ref{fig:correlation_interface} (d)). Therefore, we expect that, despite some similarity to the constructions presented in these references, our system does not have to fulfill the ``top-down'' predictions, made for gapped interfaces.

We note that even if the interface is indeed gapless, there is no contradiction here. The charge conservation rule \eqref{eq:intcn1} does not imply that the interface must be gapped. It merely makes the presence of gapping interactions possible.  Since we do not have a Hamiltonian, we do not have information on which interactions generate our wavefunctions and cannot compare them with Refs.\  \cite{cano2015interactions,maymann2019families, fliss2017interface,santos2017parafermionic, santos2018symmetry}.

%
\begin{figure}
\includegraphics[width=0.5\textwidth]{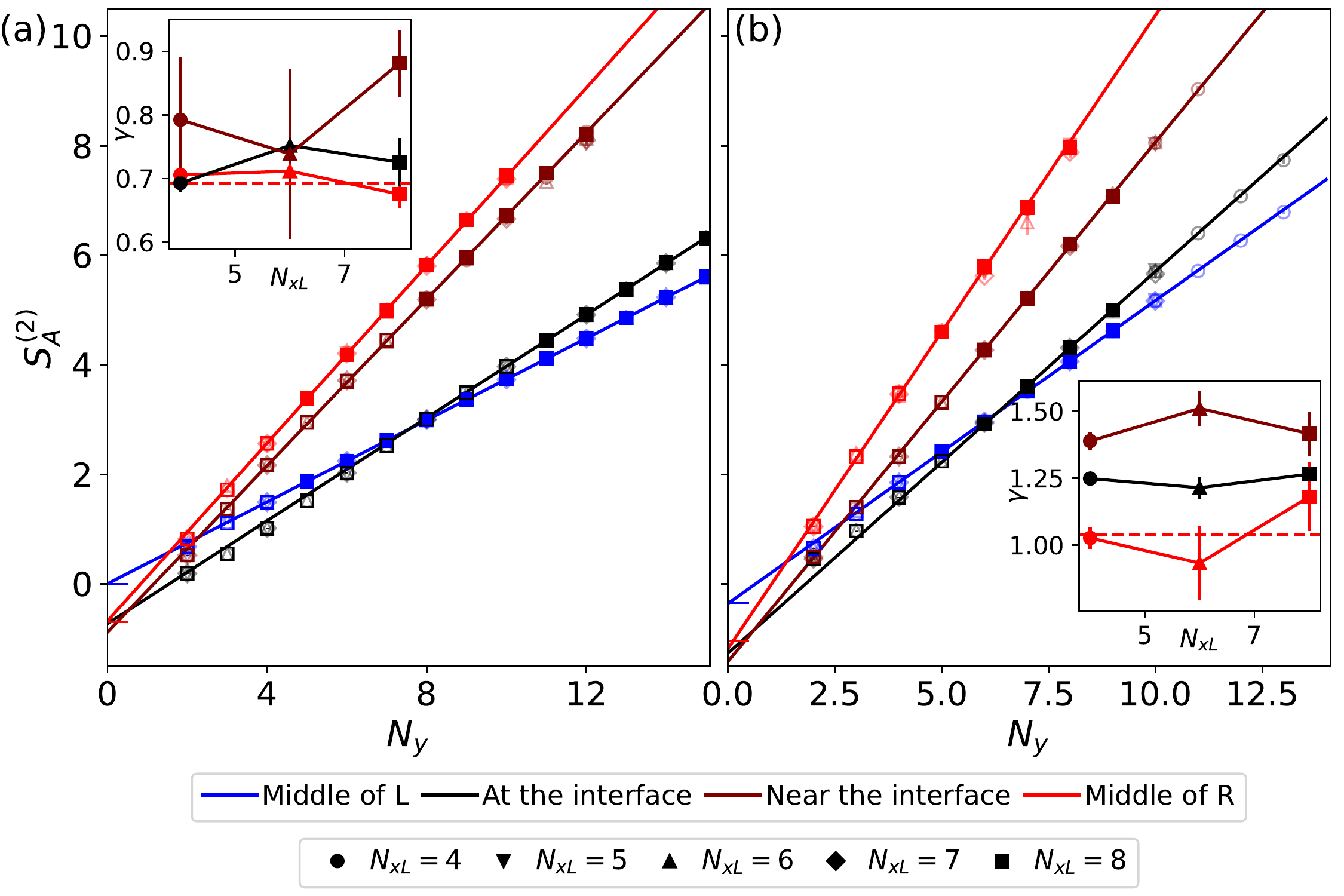}
\caption{
Scaling of the R\'enyi entropy $S^{(2)}_A$ as a function of the cylinder circumference for two considered types of systems: (a) $q_L=1, a=2$ and (b) $q_L=2, a=2$. The red and blue markers refer to the middle of the $L$ and $R$ part, respectively. The black points show the scaling precisely at the interface, while the brown points refer to the cut located slightly to the right of the interface, in which the $A$ subsystem contains the whole $L$ part and the leftmost column of sites ($2N_y$ sites) from the $R$ part. The fits are performed for the $N_{xL}=N_{xR}=8$ case, with the Monte Carlo errors included in the weights. The points included in the fit are denoted by filled markers. The theoretical values of $-\gamma_L$, $-\gamma_R$ are indicated by blue and red ticks on the $y$ axis, respectively. The plots contain also the entropies for smaller $N_{xL}=N_{xR}$, denoted by weaker colors and different marker shapes. Note that in (b) the largest possible $N_y$ decreases as we increase $N_{xL}=N_{xR}$. The insets show the fitted values of $\gamma$ for $N_{xL}=N_{xR}=4,6,8$. In (a), all the fits are performed using the same sets of $N_y$ values as in the main (a) subfigure. For (b), the details of the fits are presented in Appendix \ref{app:scaling}.
}
\label{fig:entropy}
\end{figure}

\subsection{Entanglement at the interface}\label{ssec:interface_ent}

The entanglement in the presence of an interface was studied by several authors \cite{crepel2019microscopic, crepel2019model, santos2018symmetry, cano2015interactions, bais2010topological, fliss2017interface,sakai2008entanglement}. In such a case, one can consider different entanglement cuts. For example, certain cuts crossing the interface allow to study the properties of gapless interface modes by comparing the entanglement entropy with the predictions for a 1D conformal field theory \cite{crepel2019microscopic, crepel2019model} (the validity of such an approach was confirmed analytically for the case of a single integer quantum Hall edge \cite{estienne2019entanglement}). Alternatively, a cut may be coinciding with the interface. For some gapless interfaces, numerical computations show the existence of an entanglement area law for such a cut \cite{crepel2019microscopic, crepel2019model}. The top-down works have also shown analytically that the area law exists in the case of gapped interfaces between Abelian states \cite{cano2015interactions,santos2018symmetry,fliss2017interface}. In such a case, the constant term in the linear scaling depends not only on the phases involved, but also on the interaction across the interface. This is a result of restrictions on the anyon motion, which lower the entanglement between the two parts of the system and thus increases the constant term $\gamma_{LR}$ at the interface \cite{cano2015interactions}. In particular, for the $b=1$ case, the constant term $\gamma_{LR}=\gamma_R$, where $\gamma_R$ is the topological entanglement entropy of the $R$ Laughlin state (i.e. there is a correction to the constant term only with respect to the $L$ topological entanglement entropy $\gamma_L$). Such a correction can be interpreted as originating from a symmetry-protected topological phase living at the entanglement cut (i.e. the interface) \cite{santos2018symmetry,zou2016spurious}, and thus it is connected with the existence of parafermionic modes \cite{santos2017parafermionic}. We note that the constant terms $\gamma_L$, $\gamma_R$ for the bulks of the two respective topological phases have the interpretation of topological entanglement entropy, as they are defined in a way independent from smooth deformation of the cut \cite{kitaev2006topological, levin2006detecting}. However, for $\gamma_{LR}$ such a definition does not exist, as moving the cut slightly from the interface can change $\gamma_{LR}$. 

Let us now compare predictions from Refs.\  \cite{cano2015interactions,santos2018symmetry,fliss2017interface} to Monte Carlo results for our interface wavefunction (which, as we noted in Sec. \ref{ssec:correlation}, is not necessarily gapped, so it may behave differently). We start from the $q_L=1, a=2$ case. We again cut the system parallel to the $y$ direction and investigate the scaling of the R\'enyi entropy for different $x$ positions of the cut. The results for specific positions of the cut are shown in Fig.\ \ref{fig:entropy} (a). We focus on systems with $N_{xL}=N_{xR}=8$, shown in Fig.\ \ref{fig:entropy} (a) using markers with strong colors.
Performing the linear fit \eqref{eq:fit} for the cuts in the middle of the $L$ and $R$ parts (red and blue straight lines, respectively), we can see that the entropy behaves similarly to the case of single Laughlin states (Fig.\ \ref{fig:onestate}). The fitted $\gamma_L$, $\gamma_R$ are close to theoretical values $\gamma_I=\ln(q_I)/2$ for single Laughlin states. This indicates that our wavefunction reproduces the topological orders of the $L$ and $R$ parts correctly. We note that in general the error in the entropy increases with the entropy itself. In Fig.\ \ref{fig:entropy}, we neglected the points with high error at large $N_y$, and thus there are less data points for the $R$ part than for the $L$ part.

At the interface (black markers in Fig.\ \ref{fig:entropy} (a)) the situation is more complicated. The scaling looks linear at a first glance. However, there are some deviations from the exact linear dependence. If we perform the fit using only the five largest values of $N_y$ (i.e. $N_y>10$; the black line in \ref{fig:entropy} (a))  then we obtain $\gamma_{LR}=-0.73\pm 0.04$, which is close to $\gamma_R$. This is consistent with the $\gamma_{LR}=\gamma_R$ prediction from Refs.\  \cite{cano2015interactions,santos2018symmetry,fliss2017interface}. However, one can see that the data points for low $N_y$, in particular $N_y=3$ and $N_y=4$, lie beneath the fit line. Thus, including the datapoints with lower $N_y$ would increase the fitted value of  $\gamma_{LR}$. We cannot guarantee that such departures from linear scaling do not ocurr above $N_y=15$, but if they do not, then our interface has similar scaling of entanglement entropy as the gapped interface from Refs.\  \cite{cano2015interactions,santos2018symmetry,fliss2017interface}.

To check whether the investigated scaling of the interface entropy is influenced by finite-size effects related to the finite extent of the system in the $x$ direction, we investigate systems of different $x$ size. The results for these systems are shown in Fig.\ \ref{fig:entropy} (a) using weaker colors and different marker shapes. For the middle of the $L$ and the $R$ parts, as well as for the interface, these data points are almost indistinguishable from the $N_{xL}=N_{xR}=8$, showing that these effects are negligible. To further confirm this, we perform the linear fit for $N_y>10$ and $N_{xL}=N_{xR}=4,6$ (for odd values $N_y$ can be only even, so we do not obtain enough $N_y>10$ data points). The results are shown in the inset of Fig.\ \ref{fig:entropy} (a)). The solid red and black curves, corresponding to $\gamma_R$ and $\gamma_{LR}$ both lie close to the dashed red line, which denotes the theoretical value of $\gamma_R$.

If we move our cut from the interface to the next available position to the left (so that the $A$ subsystem contains $N_y(3N_{xL}-1)$ sites), we obtain a scaling similar to the one in the bulk of the $L$ region. On the other hand, if we move it to the next available position to the right of the interface, we get a result different both from the $R$ bulk and the interface (brown markers and line in Fig.\ \ref{fig:entropy} (a)). In this case, the constant term $\gamma_{LR}'$ of the entropy scaling varies with $N_{xL}=N_{xR}$ (see the brown markers in Fig.\ \ref{fig:entropy} (a)). For $N_{xL}=N_{xR}=6$, $\gamma_{LR}'$ is close to $\gamma_R$, although with a large fit error. For $N_{xL}=N_{xR}=4,8$, $\gamma_{LR}'$ is visibly larger than $\gamma_R$. Based on our data, we are unable to determine whether this increase of $\gamma$ next to the interface is a finite-size effect, or will persist in the thermodynamic limit. Moving the cut further to the right results in a bulk-like scaling of entanglement entropy.

Similar results are obtained for $q_L=2$, $a=2$, however the picture is more distorted due to the larger correlation lengths, as well as larger values of entropy which limit the maximum available $N_y$. Here, we also obtain $\gamma_{LR}$ close to $\gamma_{R}$, although slightly larger. On the other hand, $\gamma_{LR}'$ is visibly larger than $\gamma_{R}$ for all $N_{xL}=N_{xR}$. However, the values of these terms in general depend on the points included in the fit, which shows that the finite size effects are quite pronounced for this type of the interface. More details on the $q_L=2$, $a=2$ entanglement entropy caclulations can be found in Appendix \ref{app:scaling}.

In summary, our results hint at the presence of entanglement area law at the interface and in its vicinity, as predicted in Refs.\ \cite{santos2018symmetry, cano2015interactions, fliss2017interface} for gapped interfaces and shown numerically in Refs.\ \cite{crepel2019microscopic,crepel2019model} for a gapless one. Even though the correlation function suggests gaplessness of our interface, the results regarding the entanglement entropy can be interpreted consistently with the $\gamma_{LR}=\gamma_R$ predictions from Refs.\ \cite{santos2018symmetry, cano2015interactions, fliss2017interface}. However, due to the limited size of the system and the presence of finite-size effects, we cannot guarantee that the area law holds for larger system sizes, and that $\gamma_{LR}=\gamma_R$ in the thermodynamic limit. Also, our results indicate that next to the interface, on its right side, the constant term in the entropy scaling is larger than on the interface itself. 

\section{Anyonic excitations}\label{sec:anyons}
Since both sides of the interface are topologically ordered, they have fractionalized excitations. The bottom-up approach allows us to perform detailed studies of the anyons, regarding both the universal and non-universal quantities. For single FQH states, such methods were employed to calculate the size of the anyons, their density profile, as well as to explicitly simulate their braiding \cite{nielsen2018quasielectrons,kapit2012nonabelian,liu2015characterization,johri2014quasiholes,zaletel2012exact,prodan2009mapping,
 storni2011localized,toke2007nature,jaworowski2019characterization,barbaran2009numerical,wu2015matrix, wu2014braiding, kjall2018matrix, manna2018nonabelian, manna2019quasielectrons}. However, microscopic studies of localized bulk anyonic excitations were not performed for FQH interfaces (even though it is technically possible in the MPS approach \cite{zaletel2012exact, wu2015matrix,kjall2018matrix}). Here, we fill in this gap by constructing the model wavefunctions for such excitations (Sec. \ref{ssec:anyons_wfn}), studying their charge and density profile as they cross the interface (Sec. \ref{ssec:anyons_dens}), and evaluating their statistics (Sec.  \ref{ssec:anyons_stat}).

According to Refs.\ \cite{cano2015interactions, santos2018symmetry,fliss2017interface}, there is a connection between the properties of anyons and the interface entanglement entropy scaling. For the cases studied in these papers, the motion of certain anyons through the interface is restricted, which resulted in lowering the entanglement between the $L$ and $R$ parts and the increase of $\gamma$ at the interface (leading to $\gamma_{LR}=\gamma_{R}$ in the gapped Laughlin case). In Sec. \ref{ssec:anyons_stat} we show that an analogous restriction happens in our case: statistics of some anyons become ill-defined when they cross the interface. Thus, by analogy with Refs.\  \cite{cano2015interactions, santos2018symmetry,fliss2017interface}, we regard the restriction of anyon motion as a possible explanation of $\gamma_{LR}\approx \gamma_{R}$ suggested by our results.

We also note that the connection between $\gamma$ and anyonic properties is different at the interface than in a single topological phase. In the latter case, $\gamma$ can be interpreted as topological entanglement entropy, related to the quantum dimension of the anyons. However, in the case of the interfaces, such an interpretation is not possible, since the arguments for the topological nature of this term require deformations of the cut \cite{levin2006detecting, kitaev2006topological}, which are not possible, since the interface is a 1D object. Ref.\ \cite{cano2015interactions} specifically states that the effective $K$-matrix they use to derive the entanglement correction does not describe the properties of anyons.

\subsection{Wavefunction with quasiholes}\label{ssec:anyons_wfn}
The model states for systems with quasiholes can be achieved by inserting further vertex operators into the correlator \eqref{eq:correlator}, each one corresponding to one quasihole \cite{glasser2016lattice,moore1991nonabelions}. Here, there will be two types of such operators, corresponding to two types of quasiholes: the left and right ones. The wavefunction coefficients are given by the following correlator
\begin{multline}
|\psi(\mathbf{n})|^2\propto \bra{0}\prod_{i=1}^{Q_L} V_{\mathrm{a}L}(p_{i;L}, w_{i;L}, \bar{w}_{i;L})\times \\ \times
 \prod_{i=1}^{Q_R} V_{\mathrm{a}R}(p_{i;R}, w_{i;R}, \bar{w}_{i;R})  
 \prod_{i=1}^N V(n_i, z_i, \bar{z}_i) \ket{0},
\end{multline}
where $Q_I$ is the number of quasiholes of the given type  ($I\in\{L,R \}$), $V(n_i, z_i, \bar{z}_i)$ is of the same form as in Section \ref{ssec:intwfn}, and 
\begin{equation}
V_{\mathrm{a}I}(p_{i;I}, w_{i;I}, \bar{w}_{i;I})=\normord{\exp(i \frac{p_{i;I}}{\sqrt{q_{I}}}\phi(w_{i;I},\bar{w}_{i;I}))}
\end{equation}
are the quasihole vertex operators, with $w_{i;I}$ being the positions of quasiholes and $p_{i;I}$ being integers describing their charges (in analogy to a single Laughlin state, we expect that the quasihole charge is $p_{i;I}/q_{I}$ times the charge of an $I$-type particle). The quasihole coordinates $w_{i;I}$ are external parameters of the wavefunction. The quasiholes can be located anywhere on the plane, but in this work we will put them in the middle of the smallest triangles of the kagome lattice.

Since the new vertex operators have a form analogous to \eqref{eq:vertex} (only with quasihole positions instead of particle ones), we can repeat the reasoning from Sec. \ref{ssec:correlator}, and obtain the wavefunction  
\begin{equation}
\psi(\mathbf{n})\propto \delta_{\mathbf{n}_L, \mathbf{n}_R}\psi_L(\mathbf{n}_L, \mathbf{w}_L)\psi_R(\mathbf{n}_R, \mathbf{w}_R)\psi_{LR}(\mathbf{n}, \mathbf{w})
\label{eq:wfn_anyons}
\end{equation}
where $\mathbf{w}$ is the collective label for all quasihole positions, and $\mathbf{w}_{I}$, $I\in\{L,R \}$, contains all the positions of the quasiholes of a given type. The wavefunction parts are given by
\begin{multline}
\psi_L(\mathbf{n}_L, \mathbf{w}_L)=
\prod_{i,j}(w_{i;L}-z_{j;L})^{p_{i;L}n_{j;L}}\times \\ \times
\prod_{i<j}(z_{i;L}-z_{j;L})^{q_Ln_{i;L}n_{j;L}}
\prod_{i\neq j}(z_{i;L}-z_{j;L})^{-n_{i;L}\eta_{L}},
\label{eq:wfn_anyonsL}
\end{multline}
\begin{multline}
\psi_R(\mathbf{n}_R, \mathbf{w}_R)=
\prod_{i,j}(w_{i;R}-z_{j;R})^{p_{i;R}n_{j;R}} \times \\ \times
\prod_{i<j}(z_{i;R}-z_{j;R})^{q_Rn_{i;R}n_{j;R}} 
\prod_{i\neq j}(z_{i;R}-z_{j;R})^{-n_{i;R}\eta_{R}},
\label{eq:wfn_anyonsR}
\end{multline}
\begin{multline}
\psi_{LR}(\mathbf{n}, \mathbf{w})=
\prod_{i,j}(w_{i;L}-z_{j;R})^{p_{i;L}n_{j;R}a} \times \\ \times
\prod_{i,j}(w_{i;R}-z_{j;L})^{p_{i;R}n_{j;L}/a} 
\prod_{i,j}(z_{i;L}-z_{j;R})^{aq_{L}n_{i;L}n_{j;R}} \times \\ \times
\prod_{i,j}(z_{i;L}-z_{j;R})^{-n_{i;L}\eta_{R}/a-n_{i;R}\eta_{L}a},
\label{eq:wfn_anyonsLR}
\end{multline}
\begin{multline}
\delta_{\mathrm{n}_L, \mathrm{n}_R}=\delta\Big(
q_L \left(M_L+a M_R\right) +\\+
\sum_{i=1}^{Q_L}p_{i;L}+\sum_{i=1}^{Q_R}p_{i;R}/a-N_{\phi,L}-N_{\phi,R}/a
\Big),
\label{eq:intcn3}
\end{multline}
where all the terms not dependent on the particle positions were absorbed into the normalization. When constructing this wavefunction, we chose the phase factors $\chi_i(n_i)$ to be independent from quasihole positions. Then, the calculation of quasihole statistics presented in Subsection \ref{ssec:anyons_stat} does not depend on $\chi_i(n_i)$ and thus we can set $\chi_i(n_i)=1$ for simplicity.

We note that similarly to the particles, the different types of quasiholes have different charges: an $L$-type quasihole can be replaced for example by $a$ $R$-type quasiholes with the same $p$ or one $R$-type quasihole with $a$ times larger $p$. In fact, a $p_{i;R}$ $R$-type quasihole  is fully equivalent to a $p_{j;L}=p_{i,R}/a$ $L$-type quasihole provided that $p_{i;R}$ is divisible by $a$ (one can verify that both are described by the same vertex operator). The relations between the quasiholes of different types are illustrated in Fig.\ \ref{fig:chargeconservation}.

The positions of the quasiholes are not restricted to the $L$/$R$ part of the system. However, if an $R$-type quasihole is not a valid topological excitation for the Laughlin filling $\nu=1/q_L$, its statistics will become ill-defined within the $L$ part, as we will show in Sec. \ref{ssec:anyons_stat}.

\begin{figure}
\includegraphics[width=0.5\textwidth]{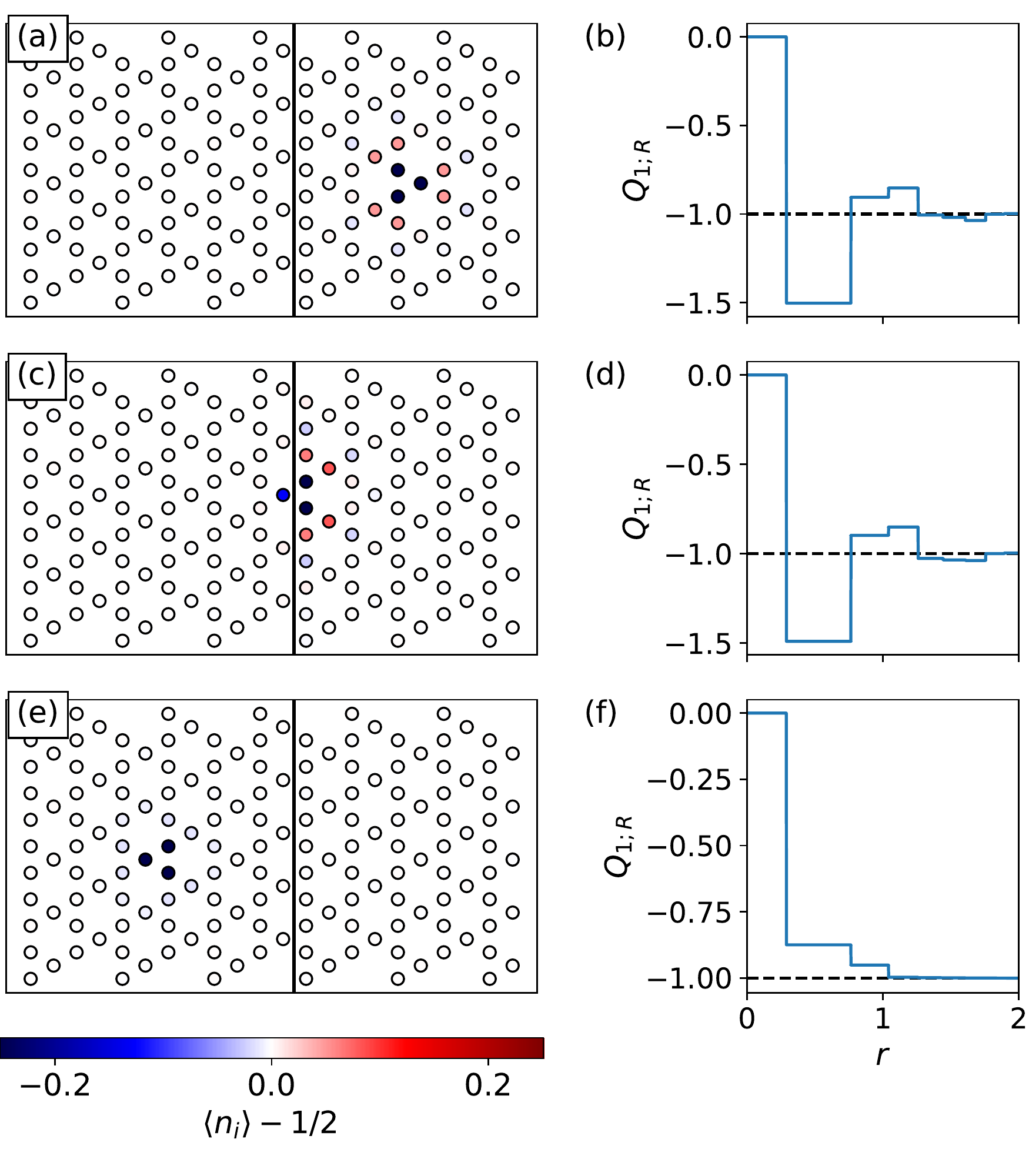}
\caption{
The density profile and charge of a $p_{1;R}=2$ $R$-type quasihole for a $q_L=1$, $a=2$ interface in a system of size $(6+5)\times 5$. The rows correspond to different positions of the quasihole: in the $R$ part (top), at the interface (middle) or in the $L$ part (bottom). The columns show different quantities: the deviation of the density distribution $\langle n_i \rangle$ from half-filling (left) and the excess charge $Q_{1;R}$ as a function of distance $r$ from the quasihole position (right). The horizontal dashed lines are located at $Q_{1;R}=-1$. Note that the distance between nearest neighbors is $r=0.5$.
}
\label{fig:anyons}
\end{figure}

\subsection{Density profile and charge of the quasiholes}\label{ssec:anyons_dens}

In the presence of a finite correlation length indicated by Fig.\ \ref{fig:onestate} (a), the quasiholes should be well localized. The Monte Carlo calculations of the particle density show that this is indeed the case. As an example, let us consider a $p_{1;R}=2$ $R$-type quasihole in a $q_L=1,a=2$ system. First, we place the quasihole in its ``parent'' part of the system, i.e. to the right of the interface (\ref{fig:anyons} (a)). The deviation from $\langle n_i \rangle=1/2$ is significant only near the quasihole position, as expected for a single FQH state. We define the excess charge within radius $r$ from the $k$-th $I$-type quasihole as 
\begin{multline}
Q_{k;I}(r)=\sum_{j}\left(\langle n_{j;L} \rangle-\frac{1}{2}\right)\theta(r- |z_{j;L}-w_{k,I}|)+\\+a\sum_{j}\left(\langle n_{j;R} \rangle-\frac{1}{2}\right)\theta(r- |z_{j;R}-w_{k,I}|),
\end{multline}
where $\theta$ is the Heaviside step function. Here, we used the fact that particles in part $L$ have unit charge, and $R$-type particles have charge $a$. The plot of the excess charge as a function of $r$ for the considered situation is shown in Fig.\ \ref{fig:anyons}(b). For large $r$ it approaches $-1$, i.e. its modulus is half the charge of an $R$ particle, as expected. 

The quasihole is well localized also when it crosses the interface, or even when it is located precisely at the border, as seen in Figs. \ref{fig:anyons}(c) and \ref{fig:anyons}(e). The corresponding excess charge plots (Figs. \ref{fig:anyons}(d) and \ref{fig:anyons}(f)) show that while the density profile of the quasihole changes, its total charge stays the same. Note that we can also interpret the charge $-1$ as a lack of one $L$-type particle, i.e. a $p_{1;L}=1$ $L$-type hole, in accordance with the charge conservation rule \eqref{eq:intcn3}. 

We observe similar behavior for quasiholes of different types and with different $p$. In principle, we can also place quasiholes in the part of the system where they are not valid topological excitations of a corresponding Laughlin state (an example is shown in Appendix \ref{app:finitesize}). In such cases, we also observe that the charge concentrated in the vicinity of the quasihole position matches our expectation for the quasihole charge. 

In some cases, moving a quasihole from the $L$ to the $R$ part or vice versa generates fluctuations of charge near the interface. This happens both for quasiholes being valid and invalid topological excitations of the given part. However, this is a finite-size effect whose strength decreases with increasing circumference of the cylinder and is expected to vanish for wide cylinders. We discuss it in Appendix \ref{app:finitesize}.

\subsection{Statistics of quasiholes}\label{ssec:anyons_stat}

Under the assumption that the quasiholes are localized, which is supported by the numerical calculations of Sec. \ref{ssec:anyons_dens}, we can obtain their statistics following the approach from Ref.\ \cite{nielsen2018quasielectrons}. We start from the wavefunction \eqref{eq:wfn_anyons} and fix the normalization constant to be real,
\begin{equation}
C=\sqrt{\sum_{\mathbf{n}_L,\mathbf{n}_R} \psi(\mathbf{n}_L,\mathbf{n}_R, \mathbf{w}_L, \mathbf{w}_R)\overline{\psi(\mathbf{n}_L,\mathbf{n}_R, \mathbf{w}_L, \mathbf{w}_R)}}
\end{equation}

The total phase in the braiding process consists of two contributions: the monodromy and the Berry phase. Let us focus on the monodromy first. Since the wavefunction contains no terms depending on the positions of two quasiholes, the only term that matters is $(w_{i;R}-z_{j;L})^{p_{i;R}n_{j;L}/a}$, for which the way the root is taken has to be defined consistently if the exponent is fractional. For a braiding process of two quasiholes in the $R$ part, $w_{i;R}$ never encircles any $L$ site (unless it goes around the cylinder -- we discuss the peculiarities of braiding particles or quasiholes in such a way in Appendix \ref{app:monodromy}). Thus, in such a case $(w_{i;R}-z_{j;L})^{p_{i;R}n_{j;L}/a}$ stays in the same branch and no phase arises from this term. On the other hand, if the braiding path contains some $L$ sites, then the contribution of $(w_{i;R}-z_{j;L})^{p_{i;R}n_{j;L}/a}$  vanishes only when $p_{i;R}$ is divisible by $a$, i.e., if the $R$-type quasihole is a vaild Laughlin anyon of the $L$ side. If not, this term yields a nonzero phase when encircling a filled $L$ site and 0 when encircling an empty one (which suggests that the mutual statistics between $L$ particles and basic $R$ quasiholes is fractional). As a consequence, the phase depends on the number of encircled $L$ particles, which is not fixed. The statistics are hence not well-defined.

Let us now proceed to the Berry phase. For concreteness, let us first assume that we move an $L$-type quasihole, whose position is denoted by $w_{1;L}$, around another quasihole, which can be of any type. The total Berry phase in the braiding process is given by
\begin{equation}
\theta=i\oint_{P}\braket{\psi|\frac{\partial}{\partial w_{1;L}}\psi}\mathrm{d}w_{1;L}+\mathrm{c.c.},
\end{equation}
where $P$ denotes the path. After inserting the wavefunction \eqref{eq:wfn_general} with coefficients \eqref{eq:wfn_anyons}, these integrals can be expressed solely in terms of the normalization constant
\begin{equation}
\theta=\frac{i}{2C^2}\oint_{P}\frac{\partial C^2}{\partial w_{1;L}}\mathrm{d}w_{1;L}+\mathrm{c.c.}
\end{equation}
By evaluating the derivative explicitly, it can be shown that the Berry phase is given by 
\begin{multline}
\theta=
\frac{i}{2}\oint_{P}\sum_{k=1}^{Q_L} \frac{p_{1;L}\langle n_{k;L}\rangle}{w_{1;L}-z_{k;A}}\mathrm{d}w_{1;L}+\\+
\frac{i}{2}\oint_{P}\sum_{k=1}^{Q_R}  \frac{ap_{1;L}\langle n_{k;R}\rangle}{w_{1;L}-z_{k;R}}\mathrm{d}w_{1;L}+ \mathrm{c.c.}
\end{multline}

To get rid of the Aharonov-Bohm phase, we subtract the phase $\theta_{\mathrm{out}}$, obtained when a second quasihole is outside the path of the first one, from the phase $\theta_{\mathrm{in}}$ obtained when the second quasihole is enclosed by the path,
\begin{multline}
\theta_{\mathrm{br}}=\theta_{\mathrm{in}}-\theta_{\mathrm{out}} =\\=
\frac{i}{2}\oint_{P}\sum_{k} \frac{p_{1;L}(\langle n_{k;L}\rangle_{\mathrm{in}}-\langle n_{k;L}\rangle_{\mathrm{out}})}{w_{1;L}-z_{k;L}}\mathrm{d}w_{1;L}+\\+
\frac{i}{2}\oint_{P}\sum_{k} \frac{ap_{1;L}(\langle n_{k;R}\rangle_{\mathrm{in}}-\langle n_{k;R}\rangle_{\mathrm{out}})}{w_{1;L}-z_{k;R}}\mathrm{d}w_{1;L} + \mathrm{c.c.}
\end{multline}
When the quasiholes are well separated, and when there is no charge accumulation on the interface, the density difference occurs only near the two positions of the second quasihole. Moreover, it does not depend on $w_{1;L}$, so it can be taken out of the integral. Applying the residue theorem, we get
\begin{multline}
\theta_{\mathrm{br}}=-2\pi p_{1;L}\sum_{k\in W_L}(\langle n_{k;L}\rangle_{\mathrm{in}}-\langle n_{k;L}\rangle_{\mathrm{out}})-\\-
2\pi a p_{1;L}\sum_{k\in W_R}(\langle n_{k;R}\rangle_{\mathrm{in}}-\langle n_{k;R}\rangle_{\mathrm{out}})
\end{multline}
where $W_I$ is the set of all $I$-type sites enclosed by the braiding path. Thus, the statistical Berry phase depends on the charge of the encircled quasihole, which, as we have shown in Sec. \ref{ssec:anyons_dens}, is constant and quantized.

For two $L$-type quasiholes, we obtain $\theta_{\mathrm{br}}=2\pi p_{1;L}p_{2;L}/q_L$, as for a single Laughlin state. For $L$ and $R$ quasiholes, the statistical Berry phase is $\theta_{\mathrm{br}}=2\pi p_{1;L}p_{1;R}/(aq_{L})$. If we repeat the derivation for moving an $R$-type quasihole, we obtain $\theta_{\mathrm{br}}=2\pi p_{1;R}p_{2;R}/q_{R}$ for encircling another $R$-type quasihole and again $\theta_{\mathrm{br}}=2\pi p_{1;L}p_{1;R}/(aq_{L})$ for encirlcing an $L$-type quasihole. Those values are the total statistical phases as long as both quasiholes are valid Laughlin anyons of the parts through which they move. If this condition is not fulfilled, the monodromy part makes the statistics ill-defined. This means that the basic $R$ quasiholes cease to be anyons as they cross the interface -- which may be interpreted as impermeability of the interface to these excitations. We also note that if the interface is gapless, the braiding whose path crosses the interface cannot be realized in an adiabatic way. Nevertheless, the mutual statistics of $L$ and $R$ quasiholes are still meaningful, as we can consider braiding around the cylinder (see Appendix \ref{app:monodromy}). Or, in planar geometry, we can envision e.g. encircling an island with filling $\nu_L=1/q_L$ embedded within a $\nu_R=1/q_R$ system.

We conclude that our interface wavefunction correctly reproduces the Laughlin quasihole statistics on each side, while introducing nontrivial statistics between the different types of quasiholes. Note that the statistics do not change when the anyons cross the interface, provided that they are well defined on both sides. Furthermore, the obtained phases are another signature of the nontrivial mutual statistics of $R$-type anyons with respect to $L$-type particles. An $L$-type quasihole of charge $p_{i;I}=q_{I}$ is equivalent to the absence of a single $L$-type particle (i.e. a hole). Thus, the statistics of an $R$-type anyon with respect to an $L$-type hole is given by $\theta_{\mathrm{br}}=-2\pi p_{1;L}/a$, which can be fractional (provided that it is well-defined, i.e. both objects are located in the $R$ part). This is a further example of the nontriviality of our interface.

\section{Conclusions}\label{sec:conclusions}

In this work, we have presented a class of model wavefunctions for interfaces between lattice Laughlin states. Our work is similar in spirit to Refs.\ \cite{crepel2019microscopic,crepel2019model, crepel2019variational}, which derived wavefunctions for the interfaces between continuum Laughlin and Halperin or Pfaffian states, with the same starting point (conformal field theory) but a different method (matrix product states). We obtained a closed-form solution similar to Laughlin's original expression \cite{laughlin}, which allowed us to calculate the properties of the system using Monte Carlo methods, sometimes aided with analytical calculations. Our work focuses on both the ground state and the localized bulk anyonic excitations. 

The study of our wavefunction yields new insights on the physics of Laughlin-Laughlin interfaces. First of all, we note that, up to our knowledge, no microscopic ansatz for the wavefunction for a Laughlin-Laughlin interface was proposed before. Our model correctly captures the topological properties of the Laughlin states on both sides, therefore it clearly describes some type of a Laughlin-Laughlin interface (although other types can exist too). 

Secondly, our system bears some similarity to the interfaces described in the ``top-down'' works and provides a microscopic realization of some of the phenomena described there. We have seen that the correct embedding of conformal field theories describing the two Laughlin states imposes a restriction on the possible filling factors \cite{bais2009condensate}. Our system realizes the charge conservation rule needed to gap out the interface \cite{cano2015interactions,maymann2019families, fliss2017interface,santos2017parafermionic, santos2018symmetry}. In addition to determining the charges of particles and anyons on both sides, we performed a microscopic simulation of an anyon crossing the interface, which was not done before. The anyon density profiles obtained from this simulation are not a topological property, but still they may yield some intuition on the anyon behavior in the general case, as for the single Laughlin state these profiles follow the same pattern in various lattice models \cite{liu2015characterization}. 

The entanglement entropy scaling at the interface can be interpreted in a way consistent with the ``top-down'' results for gapped Laughlin states  \cite{cano2015interactions, fliss2017interface, santos2018symmetry}, i.e. with the presence of area law with a constant term $\gamma_{LR}=\gamma_{R}$. It is possible that this behavior is connected with the properties of topological excitations \cite{santos2018symmetry,cano2015interactions} -- the calculation of quasihole statistics shows that some of them lose their anyonic character when they cross the interface, which can be interpreted as the impermeability of the interface to these anyons. The wavefunction gives us additional insight on the origin of this impermeability -- it arises from the monodromy and the nontrivial mutual statistics of $L$ particles and $R$ quasiholes. 

Finally, despite similarities to gapped interfaces studied in Refs.\ \cite{cano2015interactions,maymann2019families, fliss2017interface,santos2017parafermionic, santos2018symmetry}, the correlation function suggests that, if our interface can be generated with a short-range Hamiltonian, it is gapless, and thus it may be a different, less studied kind of interface. Reconciling the $K$-matrix methods from Refs.\ \cite{cano2015interactions,maymann2019families, fliss2017interface,santos2017parafermionic, santos2018symmetry,levin2013protected} with our approach would be of great interest, as it would provide additional insights on the difference between these two kinds of interfaces.

The approach taken by us has potential for further development. First, our wavefunction can be defined in more complicated geometries, such as e.g. many disconnected ``$L$'' islands within the ``$R$'' state. Defining the interface on a torus should also be possible \cite{deshpande2016lattice}. Secondly, the quasihole wavefunction can be easily generalized to quasielectrons \cite{nielsen2018quasielectrons}. Thirdly,  similar wavefunctions can be created for interfaces between Laughlin and Moore-Read states. The latter states are non-Abelian, therefore the anyon behavior is more complex. Their discretized versions have already been constructed \cite{manna2018nonabelian}. Next, one may try to generate approximate parent Hamiltonians by optimizing ground state overlaps with our wavefunctions, as it was done for single lattice quantum Hall states \cite{nielsen2013local,nandy2019truncation}. Finally, one may think about studying further exotic properties of the interface. The ``top-down'' works predict parafermionic zero modes at the ends of the gapped Laughlin-Laughlin interfaces \cite{santos2017parafermionic, santos2018symmetry}. One can wonder if such a phenomenon can be realized in our interface, and how the possible gaplessness of our interface interferes with it.

\begin{acknowledgments}
BJ was supported by Foundation for Polish Science (FNP) START fellowship no. 032.2019.
\end{acknowledgments}

\appendix

\section{Details of the $q_L=2$, $a=2$ entanglement entropy calculations}\label{app:scaling}

\begin{figure}
\includegraphics[width=0.5\textwidth]{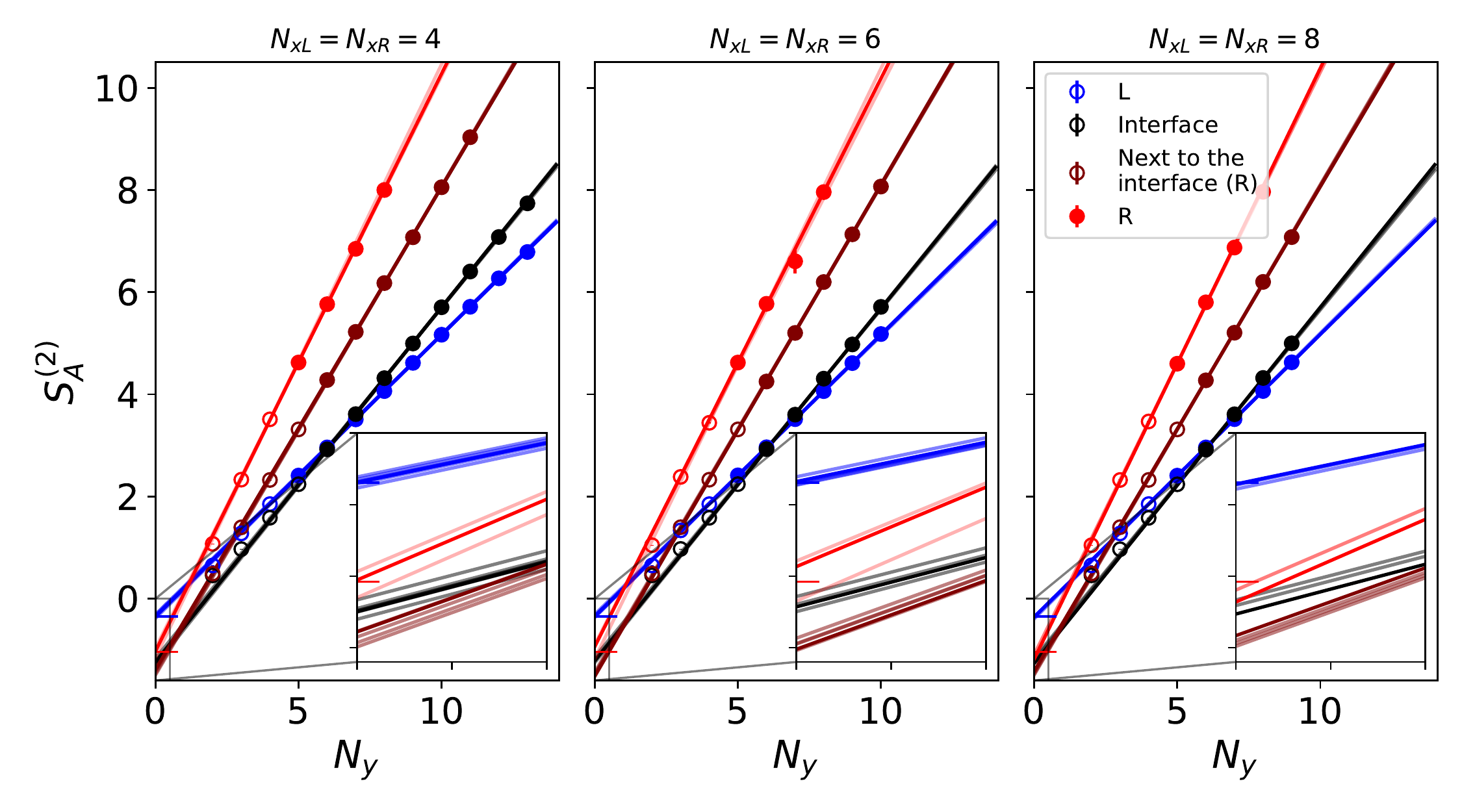}
\caption{
The entanglement entropy scaling for three different system sizes in the $x$ directions as a function of $N_y$. The blue and red markers correspond to the cut in the middle of the $L$ and $R$ parts, respectively. The black markers show the scaling precisely at the interface, while the brown points refer to the cut located slightly to the right of the interface, in which the $A$ subsystem contains the whole $L$ part and the leftmost column of sites ($2N_y$ sites) from the $R$ part. The lines with strong colors correspond to fits used in the inset of Fig.\ \ref{fig:entropy} (b). Only the data points with filled markers were included in the fit. The lines with weaker colors are alternative fits, corresponding to different numbers of points included (see the main text). The blue and red bars on the $y$ axis denote the theoretical values of the bulk $L$ and $R$ topological entanglement entropy. The insets show the magnification of the region near $N_y=0$.
}
\label{fig:qA2scaling}
\end{figure}

Since the calculations of the entanglement entropy in the $q_L=2$, $a=2$ case are affected by finite-size effects, here we provide additional data on these systems. In Fig.\ \ref{fig:qA2scaling} we plot the entanglement entropy vs. $N_y$ for three different system sizes in the $x$ direction. The lines with strong colors correspond to the fits used in the inset of Fig.\ \ref{fig:entropy} (b), with the filled markers indicating the points included in the fit. We compare them to alternative fits including points from the range $(N_{y,\mathrm{min}}, N_{y,\mathrm{max}})$. For each $N_{xL}=N_{xR}$ and each cut position, we use a fixed  $N_{y,\mathrm{max}}$ (equal to the maximum $N_y$ in Fig.\ \ref{fig:qA2scaling}) and vary $N_{y,\mathrm{min}}$ from 3 to $N_{y,\mathrm{max}}-3$ (except from the cut precisely at the interface, where the $N_y=3$ datapoint visibly departs from any linear dependence, so we neglect it and start from $N_{y,\mathrm{min}}=4$).

It can be seen that for all of the cuts except from the middle of the $L$ region, the value of $\gamma$ depends significantly on the data points included in the fit. In general, both $\gamma_{LR}$  and $\gamma_{LR}'$ are larger than $\gamma_R$. However, while $\gamma_{LR}$ remain relatively close to $\gamma_R$ (which can be interpreted as being consistent with the $\gamma_{LR}=\gamma_R$ prediction, although other interpretations are possible), $\gamma_{LR}'$ is visibly larger. Based on the data we have, we are unable to determine whether this is a finite-size effect, or $\gamma_{LR}'>\gamma_R$ persists in the thermodynamic limit (or even, whether or not the scaling is  linear for large $N_y$). 

\section{Finite-size effects from anyons crossing the interface}\label{app:finitesize}

\begin{figure}
\includegraphics[width=0.5\textwidth]{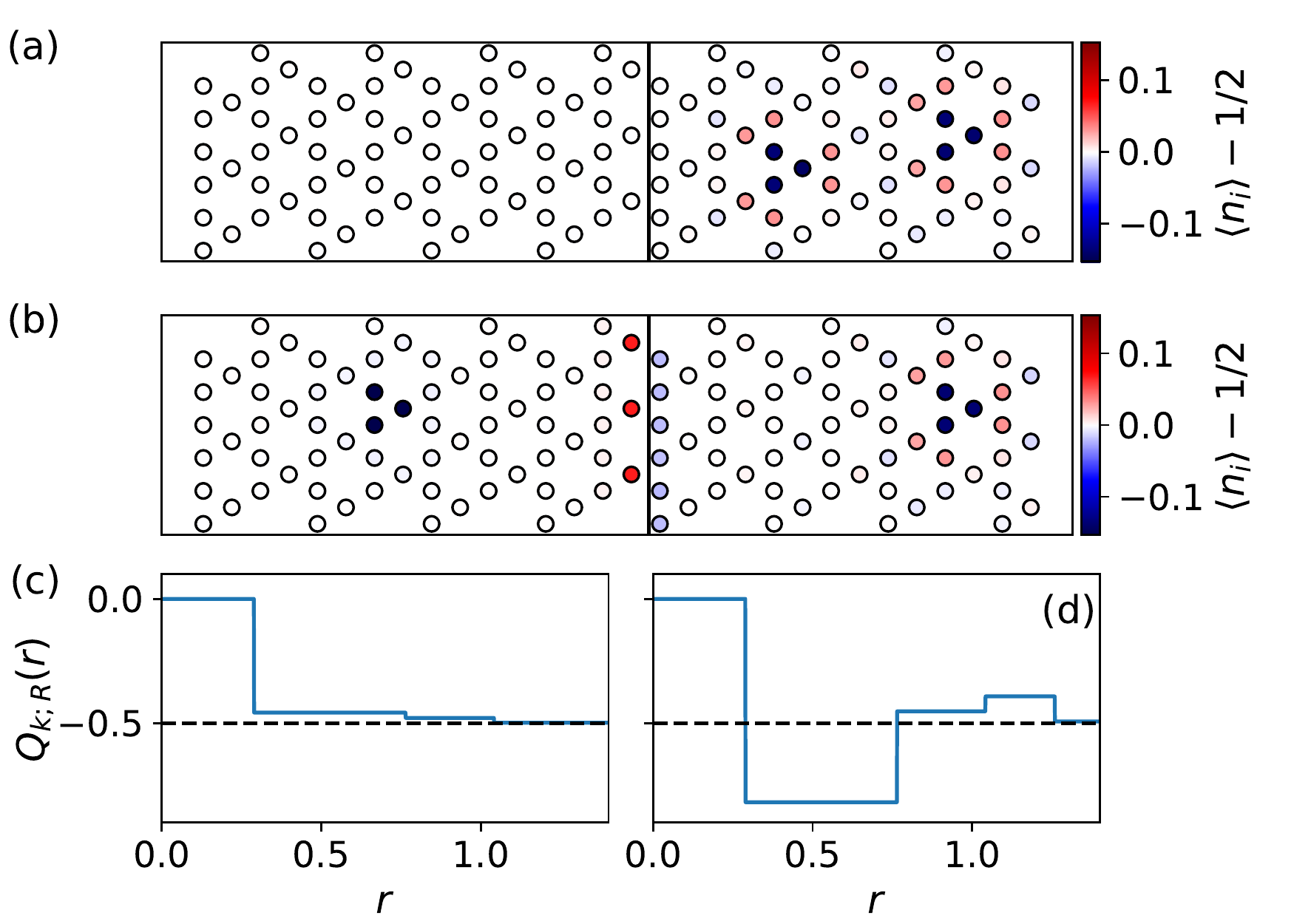}
\caption{
Two $p_{k;R}=1$ $R$-type quasiholes in the $q_L=1, a=2$ case. (a), (b) The deviation of $\langle n_i \rangle$ from 1/2 for an $(8 + 7) \times 3 $ system: (a) for both quasiholes on the $R$ side, (b) with one quasihole at each side of the interface. (c), (d) The excess charge of the quasiholes from (b), located in the $L$ and $R$ parts of the system, respectively.
}
\label{fig:finitesize1}
\end{figure}

In some cases, fluctuations of charge density occur near the interface after a quasihole is moved across it, provided that the circumference of the cylinder is small. An example for $q_L=1,a=2$  is shown in Fig.\ \ref{fig:finitesize1}. In Fig.\ \ref{fig:finitesize1} (a), two $p_{k;R}=1$ $R$-type quasiholes (each having half the charge of a basic $L$-type hole) are placed in the $R$ part, while in \ref{fig:finitesize1} (b) one of them is moved to the $L$ part. Although these quasiholes are not valid topological excitations of the $L$ part, the charge concentrated near their positions is close to $-0.5$, regardless of which part of the system they are placed in (see Fig.\ \ref{fig:finitesize1} (c), (d)). Nevertheless, the anyons cannot be regarded as fully localized, because a deviation from $\braket{n_i}=1/2$ is seen also near the interface. We stress that this does not mean that the quasihole ``leaves behind'' some of its charge at the interface when crossing it (as was predicted for gapped interfaces, e.g. in Ref.\ \cite{grosfeld2009nonabelian}), because a correct quasihole charge is observed in the vicinity of quasihole positions in Fig.\ \ref{fig:finitesize1} (b). Instead, the charge buildup probably comes from the fact that placing the quasihole to the left of the interface results in pushing some of the charge from the $L$ part towards the $R$ part, and some of this charge does not cross the interface. Note that the sum of charges accumulated on both sides of the interface vanishes.

\begin{figure}
\includegraphics[width=0.5\textwidth]{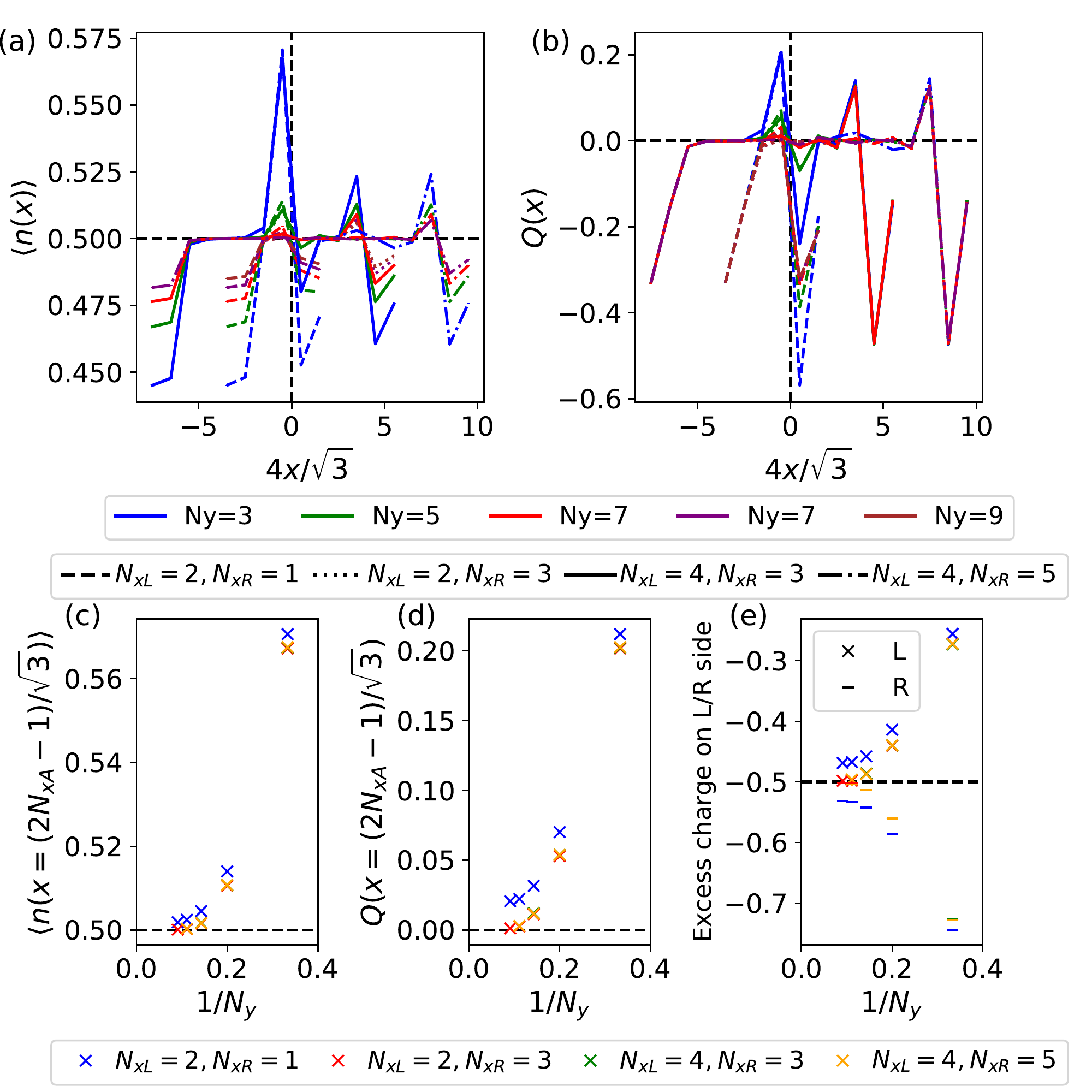}
\caption{
The accumulation of the charge at the interface in the $q_L=1, a=2$ case in systems of different size. In each case, there is a $p_{k;R}=1$ $R$-type quasihole placed near each end of the cylinder. (a) The average density as a function of the $x$ position. (b) The excess charge at a given $x$ as a function of $x$. (c) The average density at the rightmost sites of the $L$ part, as a function of $N_y$. (d) The charge accumulated at the rightmost sites of the $L$ part, as a function of $N_y$. (e) The total excess charge on each side of the interface as a function of $N_y$.  In (a) and (b), different values of $N_y$ are denoted by different colors, and different sizes in the $x$ direction by different line styles (see the legend in the middle). In (c)-(e), the different $x$ sizes are denoted by colors (see the bottom legend). 
}
\label{fig:finitesize2}
\end{figure}

This fluctuation of charge does not depend on the length of the system. This can be seen in Fig.\ \ref{fig:finitesize1} (e) depicting the average particle density at given $x$ coordinate, 
\begin{equation}
\langle n(x)\rangle =\frac{\sum_i \delta(x-x_i)\langle n_i \rangle}{\sum_i \delta(x-x_i)},
\end{equation}
for systems of different sizes. The colors denote different $N_y$s, while the line styles refer to different $N_{xL}, N_{xR}$. We focus on the density variation on the sites closest to the interface (the fluctuations occurring further from the interface are due to the presence of a quasihole at each end of the cylinder). We can observe that the density maximum on the left of the interface has similar height for different systems with the same $N_y$. The same is true for a minimum on the right of the interface (with the exception of the $N_{xL}=2$,$N_{xR}=1$, where a quasihole is too close to the interface and distorts the picture). Nevertheless, we observe that these fluctuations decrease with increasing $N_y$. This is not only an effect of averaging over more sites, as can be seen in Fig.\ \ref{fig:finitesize2} (b), showing the excess charge as a function of $x$,
\begin{multline}
Q(x) =\sum_{i=1}^{N_L} \left(\langle n_{i;L} \rangle -\frac{1}{2} \right)\delta(x-x_i)+\\
+a\sum_{i=1}^{N_R} \left(\langle n_{i;R} \rangle -\frac{1}{2} \right)\delta(x-x_i),
\end{multline}
whose variation near the interface also decreases with increasing $N_y$. Both the average density and excess charge near the interface seem to tend to their bulk values for wide cylinders (Fig.\ \ref{fig:finitesize2} (c), (d)). Moreover, the total excess charge accumulated on each side of the interface seems to converge to the charges of the quasiholes located in these regions  (Fig.\ \ref{fig:finitesize2} (e)), which means that the only excess charge is concentrated near the quasihole positions. Figs \ref{fig:finitesize2} (c)-(e) provide a further support for independence of this effect on the $x$ size of the system, as the results are very similar for all $N_{xL}, N_{xR}$ except of $N_{xL}=2$, $N_{xR}=1$.

\begin{figure}
\includegraphics[width=0.5\textwidth]{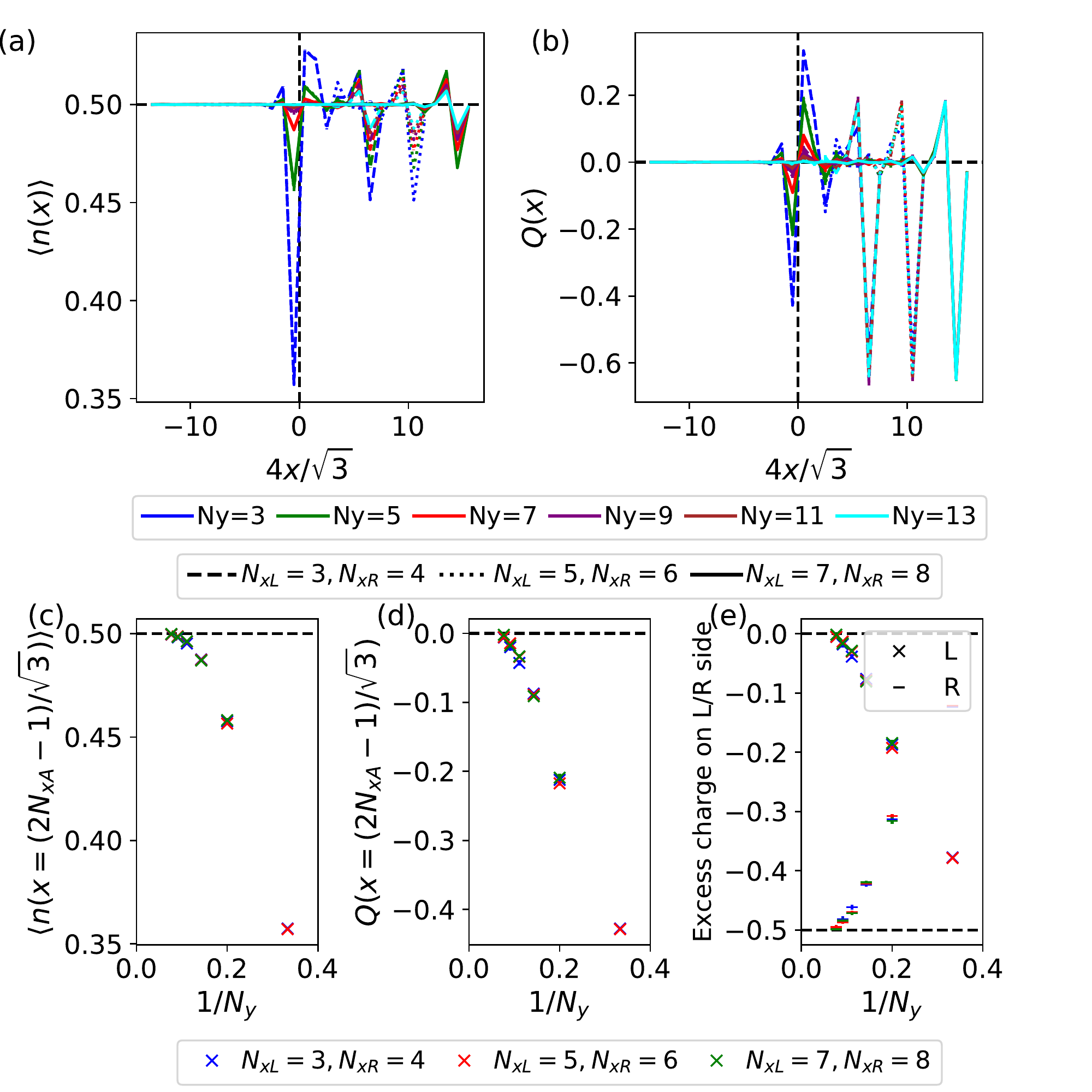}
\caption{
The accumulation of the charge at the interface in the $q_L=2, a=2$ case in systems of different size. In each case, there is a $p_{k;L}=1$ $L$-type quasihole placed near the right end of the cylinder. (a) The average density as a function of the $x$ position. (b) The excess charge at a given $x$ as a function of $x$. (c) The average density at the rightmost sites of the $L$ part, as a function of $N_y$. (d) The charge accumulated at the rightmost sites of the $L$ part, as a function of $N_y$. (e) The total excess charge on each side of the interface as a function of $N_y$.  In (a) and (b), different values of $N_y$ are denoted by different colors, and different sizes in the $x$ direction by different line styles (see the legend in the middle). In (c)-(e), the different $x$ sizes are denoted by colors (see the bottom legend). Note that the $(7+8)\times 3$ system is not taken into account, as we were not able to conduct the calculation due to large differences in the order of magnitude for the site coordinates after mapping to the complex plane.
}
\label{fig:finitesize3}
\end{figure}

Although this example above considers an $R$-type quasihole which is not a valid anyon at $\nu=1/q_L$, this is not a rule. For $q_L=2$, the charge fluctuation may occur when we move an $L$-type quasihole to the right of the interface (while all the $L$ quasiholes are valid topological excitations of the $R$ part). The example is shown in Fig.\ \ref{fig:finitesize3}. The accumulated charge decays in a way similar to the $q_L=1$ case, although more slowly, and larger size in the $x$ direction is required to separate the quasihole from the interface.

Investigating several systems of different sizes and with different quasihole positions, we observed that no charge buildup appeared when the two sides of the interface fulfilled the charge neutrality rules of the respective Laughlin states separately, as well as all the cases which can be achieved from this one by moving an equivalent of one $L$-type particle across the interface (as in Fig.\  \ref{fig:anyons}, where the quasihole has minus the charge of one $L$-type particle). In all the other cased we studied, charge fluctuations occurred at the interface if $N_y$ was small.

\section{Braiding path around the cylinder}\label{app:monodromy}

\begin{figure}
\includegraphics[width=0.5\textwidth]{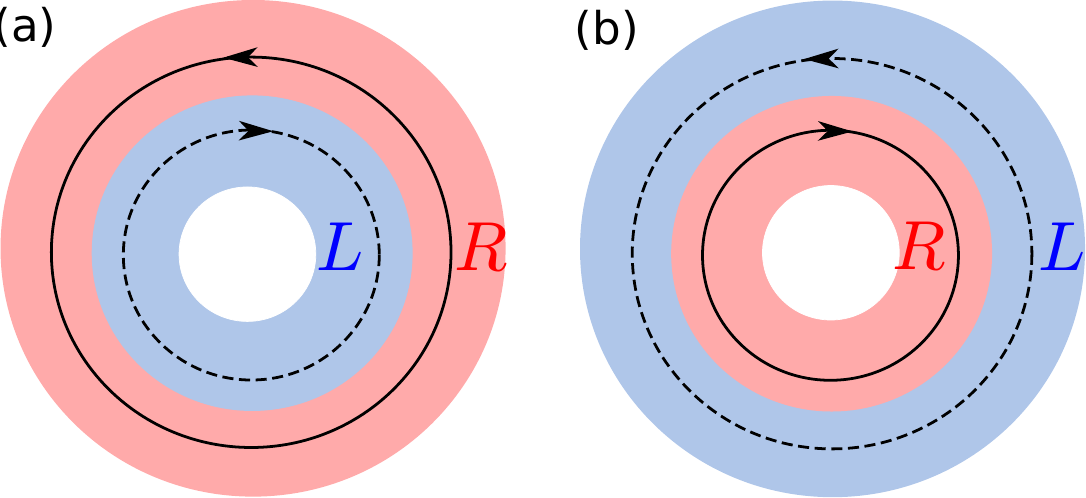}
\caption{
Two ways of mapping the cylinder to the complex plane. The circles with arrows are the paths considered in Appendix \ref{app:monodromy}: the solid one is the path of the $R$ quasihole, while the dashed one is the path of the $L$ particle.
}
\label{fig:path}
\end{figure}

The behavior of the $(w_{i;R}-z_{j;L})^{p_{i;R}n_{j;L}/a}$ term was covered in Sec.\ \ref{ssec:anyons_stat} for the paths not going around the cylinder. What happens if they do? Let us consider moving a $p_{1;R}=1$ $R$-type quasihole along a closed path which winds around the cylinder once, staying in the $R$ part throughout the process. After mapping the cylinder to the complex plane, the path looks like the solid line in Fig.\ \ref{fig:path} (a) -- it encircles the whole $L$ region. The term in question yields a phase $2\pi/a$ for each encircled filled $L$ site, i.e. $2\pi M_L/a$ in total. Now, while $M_L$ is not well-defined as the particles can be exchanged with the $R$ part, the exchange can only add or remove a multiple of $a$ $L$-type particles. Thus, the phase is in fact well-defined and equal to $\frac{2\pi}{a} (M_L \mod a)$. It does not depend on the position of the quasiholes, and thus does not contribute to statistics. 
%

There is also another potential problem raised by fractional mutual statistics of $L$ particles and $R$ quasiholes: the particle gains a nontrivial phase when encircling the quasihole, which can influence the boundary conditions for particles.  In Eqs.\ \eqref{eq:wfn_anyonsL}-\eqref{eq:wfn_anyonsLR} there are three terms which involve an $L$ particle and allow for a fractional exponent (note that $\eta_{L}$, $\eta_{R}$ can be half-integer),
\begin{multline}
\prod_{i,j}(w_{i;R}-z_{j;L})^{p_{i;R}n_{j;L}/a},~~~\prod_{i\neq j}(z_{i;L}-z_{j;L})^{-n_{i;L}\eta_{L}},\\ \prod_{i,j}(z_{i;L}-z_{j;R})^{-n_{i;L}\eta_{R}/a}.
\label{eq:terms}
\end{multline}
The first one is the term considered so far, describing the statistical phase of an $L$ particle and an $R$ quasihole, the other two describe the Aharonov-Bohm phase of an $L$ particle generated by $L$ and $R$ sites, respectively.  All these terms influence the boundary conditions for the $L$ particles.
 
In the geometry considered so far (Fig.\ \ref{fig:path} (a)), the particle-quasihole term does not raise any problems, as a path fully contained within the $L$ part (dashed line) does not encircle the $R$ part. On the other hand, we can flip the $L$ and $R$ parts before mapping the cylinder to a complex plane, which results in the geometry shown in Fig. \ref{fig:path} (b). Then, an $L$ particle can encircle the $R$ part (see the dashed line). If we consider only the $\prod_{i,j}(w_{i;R}-z_{j;L})^{p_{i;R}n_{j;L}/a}$ term, the boundary conditions for the $L$ particles seem to depend on the number of $R$ quasiholes. However, the two geometries from Fig. \ref{fig:path} describe physically equivalent systems, so if such a dependence does not exist for (a), it should not exist for (b). 

To show that this is indeed the case, we have to carefully examine all the terms generating a phase for particles \eqref{eq:terms}. We consider a path corresponding to the dashed line in Fig.\ \ref{fig:path} (b), encircling the whole $R$ part, as well as $k$ $L$ sites. We again assume that all the $R$ quasiholes with $p_{i;R}$ not divisible by $a$ are confined in the $R$ part, so that they have well-defined statistics, and hence the boundary condition for the $L$ particles is independent of the positions of these quasiholes. The total phase of the $L$ particle on the considered path is  
\begin{equation}
\phi=\frac{2\pi}{a}\left(\sum_{i=1}^{Q_R} p_{i;R}-N_{\phi;R} \right)-2\pi k\eta_L.
\end{equation}
Using the charge neutrality relation \eqref{eq:intcn3}, we obtain
\begin{multline}
\phi=-2\pi \left(q_L\left(M_L+aM_R\right) +\sum_{i=1}^{Q_L} p_{i;L}\right)+\\+
2\pi \eta_L \left(N_{L}-k\right).
\end{multline}
All the quantities in this expression except from $\eta_L$ are integers by definition, so the first term does not contribute to the phase and we are left with
\begin{equation}
\phi=2\pi \eta_L \left(N_{L}-k\right).
\end{equation}

Thus, the total phase has two interpretations: either a combination of the statistical phase of the particle with respect to the $R$ quasiholes combined with the Aharonov-Bohm phase due to the encircled $R$ sites and $k$ $L$ sites, or the Aharonov-Bohm phase due to the remaining $N_{L}-k$ sites, which are encircled when we flip the cylinder again (the dashed line in Fig.\ \ref{fig:path} (a); the difference in sign of the Aharonov-Bohm term in these two cases reflects the fact that the direction of the braiding changes when we do the flip). In other words, the boundary condition phase of the $L$ particles can be expressed using only the quantities describing the $L$ part.

\bibliography{interfaces,quasiholes}

\end{document}